\documentclass[journal,letterpaper]{IEEEtran}
\usepackage[dvips]{graphicx}
\usepackage{mathtools}
\usepackage{amsmath}
\usepackage{mathrsfs}
\usepackage{amsfonts}
\usepackage{amssymb}
\usepackage{units}
\usepackage{multirow}
\usepackage{color}
\usepackage{cite}
\usepackage{algorithm}
\usepackage{algpseudocode}

\newcommand{\data}[1]{\breve{#1}}
\newcommand{\model}[1]{{#1}}

\newcommand{\onehalf}{\nicefrac{1}{2}}

\newcommand{\mat}[1]{\mathbf{#1}} % FOR NUMERIC MATRICES
\newcommand{\vet}[1]{\mbox{\boldmath $#1$}} % FOR VECTORS (both numeric, signals, and transformed signals)
\newcommand{\tran}{\mathsf{T}}
\newcommand{\herm}{\mathsf{H}}

\newcommand{\pert}[1]{\widehat{#1}}

\newcommand{\abs}[1]{\left\lvert #1 \right\rvert }
\newcommand{\norm}[1]{ \left\lVert #1 \right \rVert}

\def\XXint#1#2#3{{\setbox0=\hbox{$#1{#2#3}{\int}$}
\vcenter{\hbox{$#2#3$}}\kern-.5\wd0}}

\def\jj{\mathrm{j}}
\renewcommand{\Re}[1]{\mathrm{Re}\left \{#1\right\} }
\renewcommand{\Im}[1]{\mathrm{Im}\left\{#1\right\}}
\newcommand{\Real}{\mathbb{R}}
\newcommand{\Complex}{\mathbb{C}}

\newcommand{\junk}[1] {}

\def\xx{\vet{x}} % Perturbation variables, vectorized
\begin{document}

% paper title
\title{A Perturbation Scheme for Passivity Verification and Enforcement of Parameterized Macromodels}

\author{Stefano~Grivet-Talocia,~\IEEEmembership{Senior~Member,~IEEE}%
\thanks{%
Submitted to the IEEE Transactions on components, packaging and manufacturing technology on 13-Apr-2017}%
\thanks{%
S.~Grivet-Talocia is with the Department of Electronics and Telecommunications, Politecnico di Torino, Torino~10129, Italy (e-mail: stefano.grivet@polito.it).}%
}

\maketitle
%%%%%%%%%%%%%%%%%%%%%%%%%%%%%%%%%%%%%%%%%%%%%%%%%%%%%%%%%%%

%%%%%%%%%%%%%%%%%%%%%%%%%%%%%%%%%%%%%%%%%%%%%%%%%%%%%%%%%%%
\begin{abstract}
This paper presents an algorithm for checking and enforcing passivity of behavioral reduced-order macromodels of LTI systems, whose frequency-domain (scattering) responses depend on external parameters. Such models, which are typically extracted from sampled input-output responses obtained from numerical solution of first-principle physical models, usually expressed as Partial Differential Equations, prove extremely useful in design flows, since they allow optimization, what-if or sensitivity analyses, and design centering.

Starting from an implicit parameterization of both poles and residues of the model, as resulting from well-known model identification schemes based on the Generalized Sanathanan-Koerner iteration, we construct a parameter-dependent Skew-Hamiltonian/Hamiltonian matrix pencil. The iterative extraction of purely imaginary eigenvalues ot fhe pencil, combined with an adaptive sampling scheme in the parameter space, is able to identify all regions in the frequency-parameter plane where local passivity violations occur. Then, a singular value perturbation scheme is setup to iteratively correct the model coefficients, until all local passivity violations are eliminated. The final result is a corrected model, which is uniformly passive throughout the parameter range. Several numerical examples denomstrate the effectiveness of the proposed approach.
\end{abstract}
%%%%%%%%%%%%%%%%%%%%%%%%%%%%%%%%%%%%%%%%%%%%%%%%%%%%%%%%%%%

%%%%%%%%%%%%%%%%%%%%%%%%%%%%%%%%%%%%%%%%%%%%%%%%%%%%%%%%%%%
%%%%%%%%%%%%%%%%%%%%%%%%%%%%%%%%%%%%%%%%%%%%%%%%%%%%%%%%%%%

%%%%%%%%%%%%%%%%%%%%%%%%%%%%%%%%%%%%%%%%%%%%%%%%%%%%%%%%%%%
\section{Introduction and Motivation}
%%%%%%%%%%%%%%%%%%%%%%%%%%%%%%%%%%%%%%%%%%%%%%%%%%%%%%%%%%%

This paper addresses the general problems of passivity verification and passivity enforcement for behavioral models of LTI systems, whose transfer function depends on external parameters. This problem arises in all those application areas that require reduced-order, compact macromodels that can be simulated efficiently either in frequency or time-domain using standard Ordinary Differential Equation (ODE) or legacy circuit solvers of the SPICE class, in order to assess the performance of a given underlying complex circuit or system~\cite{PM_book,nakhla01,jnl-2010-temc-pi,CelikBook}. Typical application scenarios include Signal and Power Integrity verification of digital, analog, mixed-signal and RF systems including all parasitic effects of electrical interconnects at chip, package and board level, reduced-order modeling of circuit blocks for speeding up simulation tasks, or Computer Aided Design via virtual prototyping, including sensitivity, what-if and design centering processes.

Behavioral reduced-order macromodels are typically derived either through model order reduction of large circuit equations cast in state-space or descriptor form~\cite{antoulas2005,schilders2008,CelikBook}, whenever such characterizations are available (e.g., through a partial element equivalent circuit extraction, or direct spatial discretization of Maxwell's equations), or through a model identification from measured or simulated port responses~\cite{PM_book,Gustavsen99,ovf2,desc2008,artAJMACA,SLACATCAD09}. Only the latter case is considered in this work, due to its widespread adoption by industry, academia, and simulation tool vendors. Therefore, we target here the construction of a passive model in state-space or descriptor form of a generic LTI multiport system starting from sampled frequency responses at its terminals.

Several approaches are available for macromodel generation in the univariate case, where the system transfer function or matrix depends only on frequency (see~\cite{PM_book} for a comprehensive review). Since model extraction is based on fitting a prescribed model structure to available data through some numerical optimization, the macromodel responses are inevitably affected by some approximation error. This error cannot be avoided, since most often the underlying system is characterized by non-rational transfer functions (e.g., in the common case of distributed systems with lossy and dispersive materials). Fitting a (finite-order) rational model to represent such systems cannot lead to exact identification. Such approximation error can be sufficient to make a model non-passive, even when it is supposed to represent a passive physical system. Lack of passivity is very critical and may lead to instabilities when using the models during system-level simulations~\cite{Triverio07,jnl-2009-ijcta-destabilize}. Therefore, any passivity violation must be carefully identified and removed.

Passivity verification and enforcement of univariate models is well-established. The common tool that provides an algebraic test for passivity is the so-called Hamiltonian matrix associated to a state-space realization of the model~\cite{Grivet04pass,BBK89}. The presence of purely imaginary eigenvalues of such matrix pinpoints passivity violations (e.g., frequency bands where the model has active behavior). In case of such violations, several model perturbation methods can be used to correct the model coefficients and recover model passivity~\cite{PM_book,jnl-2008-tadvp-PassivityMethods}. Most prominent methods are based on Hamiltonian eigenvalue perturbation~\cite{Grivet04pass,SHH_perturbation}, singular value/eigenvalue perturbation~\cite{saraswat2005,Gust01}, Positive or Bounded Real Lemma~\cite{LMI} enforcement based on semidefinite programming~\cite{silveira}, $\mathcal{H}_\infty$ norm optimization via non-smooth (convex) optimization~\cite{jnl-2012-tmtt-subgradient} or localization methods~\cite{jnl-2014-tcad-localization}.

The main objective of this work is the extension of the above passivity verification and enforcement methods to the multivariate case, where the transfer function of the model depends not only on frequency, but also on additional external parameters. Examples can be geometrical or material properties of the underlying physical circuit or system~\cite{TriverioPhD,jnl-2007-temc-Parameterization,Triverio10,Triverio07EPEP,Ferranti10b}, temperature or biasing conditions~\cite{jnl-2014-tcpmt-small-signal}, etc.

Few approaches are able to provide parameterized models that are uniformly passive througout the parameter domain. These approaches impose some constraints on the model structure and/or the model identification procedure. For instance, we can cite the interpolatory approaches that first construct univariate models for fixed parameter values, and then recover a multivariate model by interpolating such ``grid'' or ``root'' macromodels. If the root macromodels are passive and if a passivity-preserving interpolation scheme is used, also the multivariate model will be uniformly passive~\cite{Samuel13,Ferranti12b,Ferranti11,Ferranti10b}. This approach has two main problems. First, passivity-preserving interpolation schemes are overconstrained and may lead to inaccurate behavior of the interpolant. This problem is usually addressed by increasing the number of root macromodels, thus degrading the efficiency of the identification process. Second, the multivariate model is cast as a linear combination of root macromodels. Therefore, if each root macromodel is characterized by a given order $\bar{n}$, the multivariate model will have a larger order $L\bar{n}$, where $L$ is the number of root macromodels contributing to the interpolation. On one hand, this model structure may not be related to the physics of the underlying system (if $\bar{n}$ natural frequencies are required, this should be true for all parameter values). On the other hand, model complexity is larger than strictly necessary.

The approach that we follow in this work is different. We embed the parameter dependence through a dedicated model structure, based on a suitable expansion into frequency-dependent and parameter-dependent basis functions~\cite{Triverio2009,Triverio10,Triverio07EPEP,jnl-2014-tcpmt-small-signal}. Thus, model structure is general and can be tuned by selecting the appropriate basis functions for the specific application at hand. Model coefficients are then identified through a Generalized Sanathanan-Koerner iteration~\cite{San63}. This approach is standard~\cite{TriverioPhD,Triverio2009}.

The novel contribution of this work is twofold. First, we devise an adaptive sampling scheme in the parameter space based on some special features of the Hamiltonian eigenspectrum. This scheme is able to pinpoint all passivity violations throughout the frequency and parameter plane. Second, we extend a known passivity enforcement scheme for univariate models based on singular value perturbation to the multivariate case. The result is an iterative passivity enforcement scheme for multivariate macromodels that is able to guarantee uniform passivity as well as model accuracy with constant model order throughout the parameter space. To the best of Authors' knowledge, this is the first passivity verification and enforcement scheme that is applicable to multivariate models. We focus this work on the case of a single external parameter, or equivalently on bivariate macromodels. The extension to the muldimensional case will be documented in a future report.

%%%%%%%%%%%%%%%%%%%%%%%%%%%%%%%%%%%%%%%%%%%%%%%%%%%%%%%%%%%
\section{Preliminaries and Notation}\label{sec:notation}
%%%%%%%%%%%%%%%%%%%%%%%%%%%%%%%%%%%%%%%%%%%%%%%%%%%%%%%%%%%

Let us consider a generic LTI multiport structure with transfer matrix $\data{\mat{H}}(s;\vartheta)$, where $s$ is the Laplace variable, and $\vartheta \in \Theta=[\vartheta_{\min},\vartheta_{\max}]$ is some external parameter, e.g., some material or geometrical characteristic of the underlying physical system. Throughout this work, it is assumed that $\data{\mat{H}}(s;\vartheta) \in \mathbb{C}^{P\times P}$ is the scattering matrix of the system at well-defined ports. The accent $\,\data{}\,$ will be used to label the original or ``true'' system responses, that are generally unknown.

The ``true'' response is assumed to be available, either via direct measurements or via numerical calculation based on a first-principle model (e.g., the numerical solution of Maxwell's equations for an electromagnetic system) at a set of fixed frequencies $f_k,\;k=1,\dots,\bar{k}$ over a given frequency band $[f_{\min},f_{\max}]$, with $f_1=f_{\min}$ and $f_{\bar{k}}=f_{\max}$, and for a set of fixed parameter values $\vartheta_m,\;m=1,\dots,\bar{m}$ spanning the range $[\vartheta_{\min},\vartheta_{\max}]$. We will denote this characterization as
\begin{equation}\label{eq:data}
	\data{\mat{H}}_{k,m} = \data{\mat{H}}(\jj 2\pi f_k; \vartheta_m),\quad k=1,\dots,\bar{k},\; m=1,\dots,\bar{m}.
\end{equation}

Since a closed-form dependence of the true system response on $s$ and $\vartheta$ is not available, there comes the need of constructing a parameterized model that fits or interpolates the above data points, and that can be solved efficiently in frequency and time domain. We assume here the following model structure 
\begin{equation}\label{eq:gsk_structure}
	\model{\mat{H}}(s;\vartheta)  = \dfrac{\textsf{\textbf{N}}(s,\vartheta)}{\mathsf{D}(s,\vartheta)} 
		 = \dfrac{\sum_{n=0}^{\bar{n}} \mat{R}_{n}(\vartheta)\, \varphi_n(s)}%
			{\sum_{n=0}^{\bar{n}}  r_{n}(\vartheta)\, \varphi_n(s)}.
\end{equation}
The frequency-dependent basis functions $\varphi_n(s)$ are partial fractions based on a prescribed set of distinct $\bar{n}_r$ real poles $q_n\in\Real^-$ and $\bar{n}_c$ complex pole pairs $p_n=p'_n\pm \jj p''_n\in\Complex^-$, where $\varphi_0(s)=1$,
\begin{equation}
	\varphi_n(s) = (s-q_n)^{-1}, \quad \textrm{for}\quad n=1,\dots,\bar{n}_r,
\end{equation}
and
\begin{equation}
	\begin{aligned}
	\varphi_{\bar{n}_r+2\nu-1}(s) & =  (s-p_\nu)^{-1} + (s-p^*_\nu)^{-1}\\
	\varphi_{\bar{n}_r+2\nu}(s) & =  \jj (s-p_\nu)^{-1} - \jj (s-p^*_\nu)^{-1}
	\end{aligned}
\end{equation}
for $\nu = 1,\dots,\bar{n}_c$, where $^*$ denotes the complex conjugate. The (real-valued) numerator and denominator coefficients are further expressed as
\begin{equation}\label{eq:param_coeff}
	 \mat{R}_{n}(\vartheta) = \sum_{\ell=1}^{\bar{\ell}} \mat{R}_{n,\ell} \,\xi_\ell(\vartheta),\quad
	 r_{n}(\vartheta) = \sum_{\ell=1}^{\bar{\ell}} r_{n,\ell} \,\xi_\ell(\vartheta)
\end{equation}
where the basis functions $\xi_\ell(\vartheta)$ are responsible for reproducing the variations induced by the external parameter $\vartheta$. We remark that the adopted model structure is the same as discussed in~\cite{PM_book} and originally introduced in~\cite{Triverio2009,Samuel13,Ferranti11,Ferranti10b,jnl-2017-temc-fourier-rational}, where polynomials, piecewise polynomials, or trigonometric polynomials were used as basis functions $\xi_\ell$. This work makes no a-priori assumption on the specific choice of basis functions, which should be defined considering the particular application at hand.

Model structure~\eqref{eq:gsk_structure} guarantees that $\model{\mat{H}}(s;\vartheta)$ is a rational function of $s$, with implicitly parameterized poles and residues (explicit parmeterization of model poles is to be avoided, due to the possibly non-smooth behavior in case of bifurcations~\cite{jnl-2017-temc-fourier-rational}). Thanks to the assumed linear dependence of both model numerator $\textsf{\textbf{N}}(s,\vartheta)$ and denominator $\mathsf{D}(s,\vartheta)$ on the respective coefficients $\mat{R}_{n,\ell}$ and $r_{n,\ell}$, the latter can be easily evaluated during a model identification step by enforcing the fitting condition
\begin{equation}\label{eq:fitting}
	\model{\mat{H}}(\jj 2\pi f_k; \vartheta_m)  \approx \data{\mat{H}}_{k,m},\quad k=1,\dots,\bar{k},\; m=1,\dots,\bar{m}.
\end{equation}
in least squares sense. This is achieved here through a linear relaxation of~\eqref{eq:fitting} known as (Generalized) Sanathanan-Koerner (GSK) iteration~\cite{PM_book,Triverio2009,San63}. This approach transforms the nonlinear least-squares problem arising from~\eqref{eq:fitting} into an iterative sequence of weighted linear least squares problem, whose solution is straightforward. This procedure is standard in parameterized model identification and is not further discussed here. The Reader is referred to~\cite{PM_book} for more details on the GSK iteration, and to~\cite{lefteriu2013,shi2016} for a discussion on its convergence properties. We finally remark that, in order to guarantee uniqueness in the model representation, we normalize the model coefficients by setting $r_{0,1}=1$.

%%%%%%%%%%%%%%%%%%%%%%%%%%%%%%%%%%%%%%%%%%%%%%%%%%%%%%%%%%%
\section{Problem statement}\label{sec:problem}
%%%%%%%%%%%%%%%%%%%%%%%%%%%%%%%%%%%%%%%%%%%%%%%%%%%%%%%%%%%

The above-mentioned GSK iteration is not able to enforce model and passivity by construction, since no explicit passivity constraints are enforced during model identification. We recall that the model is passive for a given parameter value $\vartheta$ if and only if the corresponding scattering matrix $\model{\mat{H}}(s;\vartheta)$ is Bounded Real~\cite{Wohlers,Anderson}:
\begin{enumerate}
	\item $\model{\mat{H}}(s;\vartheta)$ is regular for $\Re{s} > 0$,
	\item $\model{\mat{H}}^*(s;\vartheta) = \model{\mat{H}}(s^*;\vartheta)$,
	\item $\mat{I}_P - \model{\mat{H}}^\herm(s;\vartheta)\model{\mat{H}}(s;\vartheta)\geq 0$ for $\Re{s} > 0$,
\end{enumerate}
where $^\herm$ is the Hermitian transpose, and $\mat{I}_P$ is the identity matrix of size $P$. Condition 1) is related to (asymptotic) stability, which is here assumed a priori (as can be readily veryfied by a suitable parameter sweep of the parameter-dependent model poles), whereas condition 2) ensuring a real impulse response is automatically satisfied thanks to the assumed model structure~\eqref{eq:gsk_structure}. Condition 3), implying no energy gain from the model throughout the open right complex plane, is instead more difficult to check and to enforce. Note that it is sufficient to check condition 3) only on the imaginary axis $s=\jj\omega$,
\begin{equation}\label{eq:BR_cond_3}
	\mat{I}_P - \model{\mat{H}}^\herm(\jj\omega;\vartheta)\model{\mat{H}}(\jj\omega;\vartheta)\geq 0\quad \forall\omega\in\mathbb{R},
\end{equation}
which in turn is equivalent to
\begin{equation}\label{eq:BR_cond}
	\sigma_{\max} \{ \model{\mat{H}}(\jj\omega;\vartheta)  \}   \leq 1, \quad \forall\omega\in\mathbb{R},
\end{equation}
where $\sigma_{\max}\{\cdot\}$ extracts the largest singular value of its matrix argument.

Several solutions are in fact available for checking and enforcing~\eqref{eq:BR_cond} on univariate models that depend only on frequency, which in our setting would correspond to the model instantiated for a fixed value of the external parameter $\vartheta$. For a comprehensive review of such methods, the Reader is referred to~\cite{PM_book} and to~\cite{jnl-2008-tadvp-PassivityMethods}. What we are interested in this work is the uniform passivity of the parameterized model $\model{\mat{H}}(s;\vartheta)$ throughout the parameter range, so that the above Boudned Realness conditions, in particular~\eqref{eq:BR_cond}, will hold $\forall\vartheta \in [\vartheta_{\min},\vartheta_{\max}]$.

The approach that is pursued here can be regarded as an extension to the multivariate case, of existing standard approaches valid for non-parameterized systems. We start from some initial non-passive model $\model{\mat{H}}(s;\vartheta)$, and we perturb its numerator coefficients as
\begin{equation}\label{eq:pert_coefficients}
	\pert{\mat{R}}_{n,\ell} = \mat{R}_{n,\ell} + \Delta\mat{R}_{n,\ell} \quad \forall n,\ell
\end{equation}
by determining the correction $\Delta\mat{R}_{n,\ell}$ that is required to ensure that the perturbed scattering response of the parameterized model $\pert{\model{\mat{H}}}(s;\vartheta)$ is uniformly Bounded Real in the prescribed parameter range. Note that we do not perturb the denominator coefficients, since we would like to preserve the model poles, i.e., the zeros of the denominator $\mathsf{D}(s,\vartheta)$. This choice is standard in practically all passivity enforcement methods applicable to univariate models.

Setting up the model perturbation requires a precise localization of the regions in the frequency-parameter plane where the Bounded Realness condition~\eqref{eq:BR_cond} is violated. This problem is addressed in Section~\ref{sec:check}, which first casts some known results for univariate models in the parameterized framework (Section~\ref{sec:check_univariate}) and then extends the check to the multivariate case (Section~\ref{sec:check_multivariate}). Section~\ref{sec:enforcement} presents the main passivity enforcement scheme. In particular, Section~\ref{sec:constraints} constructs the algebraic constraints that, when iteratively enforced, lead to removal of all passivity violations; Section~\ref{sec:cost} constructs a cost function whose minimization will ensure preservation of model accuracy during perturbation; and Section~\ref{sec:iterations} presents the main iterative passivity enforcement scheme. All these developments require the construction of a parameter-dependent descriptor realization of the model, which is discussed next.

%%%%%%%%%%%%%%%%%%%%%%%%%%%%%%%%%%%%%%%%%%%%%%%%%%%%%%%%%%%
\section{Descriptor Realizations}\label{sec:descriptor}
%%%%%%%%%%%%%%%%%%%%%%%%%%%%%%%%%%%%%%%%%%%%%%%%%%%%%%%%%%%

Starting from model~\eqref{eq:gsk_structure}, we define the following descriptor realization
\begin{equation}\label{eq:descriptor}
\left\{
	\begin{aligned}
	\mat{E} \dot{\vet{x}} & =  \mat{A}(\vartheta) \vet{x} + \mat{B} \vet{u} \\
	\vet{y} & =  \mat{C}(\vartheta) \vet{x}
	\end{aligned}
\right.
\end{equation}
where $\vet{x}\in\mathbb{R}^{N+P}$ denotes internal generalized states with $N = \bar{n} P$, and $\vet{u},\vet{y}\in\mathbb{R}^{P}$ are incident and reflected scattering waves at the model ports. The descriptor matrices are constructed as
\begin{equation}\label{eq:descr_matrices}
	\begin{array}{ll}
	\mat{E} = \begin{bmatrix} \mat{I}_N & \mat{0}_{N,P} \\  \mat{0}_{P,N} & \mat{0}_{P,P} \end{bmatrix}, &
	\mat{A}(\vartheta) = \begin{bmatrix} \mat{A}_0 & \mat{B}_0 \\  \mat{C}_2(\vartheta) & \mat{D}_2(\vartheta) \end{bmatrix},  \\
	\\
	\mat{C}(\vartheta) = \begin{bmatrix}  \mat{C}_1(\vartheta) & \mat{D}_1(\vartheta) \end{bmatrix}, &
	\mat{B}(\vartheta) = \begin{bmatrix}  \mat{0}_{N,P} \\ -\mat{I}_P(\vartheta) \end{bmatrix},
	\end{array}
\end{equation}
where $\mat{0}_{J,K}$ denotes the all-zero matrix block of size $J \times K$. In~\eqref{eq:descr_matrices}, the matrices $\{\mat{A}_0, \mat{B}_0, \mat{C}_1(\vartheta), \mat{D}_1(\vartheta)\}$ and $\{\mat{A}_0, \mat{B}_0, \mat{C}_2(\vartheta), \mat{D}_2(\vartheta)\}$ provide regular state-space realizations of model numerator $\textsf{\textbf{N}}(s,\vartheta)$ and ``extended'' denominator $\mathsf{D}(s,\vartheta) \mat{I}_P$, respectively. Individual matrices in these realizations are here defined as
\begin{equation}
	\begin{aligned}
	\mat{A}_0 & = {\rm blkdiag}\{ \mat{A}_{0r}, \mat{A}_{0c} \} \\
	\mat{B}_0^\tran & = \begin{bmatrix} \mat{B}_{0r}^\tran, \mat{B}_{0c}^\tran \end{bmatrix} \\
	\end{aligned}
\end{equation}
where $^\tran$ is the matrix transpose, with
\begin{equation}
	\begin{aligned}
	\mat{A}_{0r} & = {\rm blkdiag}\{ q_n \mat{I}_P \}_{n=1}^{\bar{n}_r} \\
	\mat{A}_{0c} & = {\rm blkdiag}\left\{ \begin{bmatrix} p'_n \mat{I}_P & p''_n \mat{I}_P \\
	-p''_n \mat{I}_P & p'_n \mat{I}_P \end{bmatrix} \right\}_{n=1}^{\bar{n}_c} \\
	\mat{B}_{0r} & =  \begin{bmatrix} 1,\dots,1 \end{bmatrix}^\tran \otimes  \mat{I}_P \\
	\mat{B}_{0c} & =  \begin{bmatrix} 2,0,\dots,2,0 \end{bmatrix}^\tran \otimes  \mat{I}_P
	\end{aligned}
\end{equation}	
where $\otimes$ is the Kronecker product, and
\begin{equation}
	\begin{aligned}
	\mat{C}_1(\vartheta) & = \begin{bmatrix} \mat{R}_1(\vartheta) & \cdots &  \mat{R}_{\bar{n}}(\vartheta) \end{bmatrix} \\
	\mat{C}_2(\vartheta) & = \begin{bmatrix} \mat{I}_P \, r_1(\vartheta) & \cdots &  \mat{I}_P\,  r_{\bar{n}}(\vartheta) \end{bmatrix} \\
	\mat{D}_1(\vartheta) & = \mat{R}_0(\vartheta) \\
	\mat{D}_2(\vartheta) & = \mat{I}_P\, r_0(\vartheta)
	\end{aligned}
\end{equation}
It is straightforward to show that, with the above definitions, the transfer matrix associated to the descriptor form~\eqref{eq:descriptor} 
\begin{equation}\label{eq:descr_tf}
	\model{\mat{H}}(s;\vartheta) = \mat{C}(\vartheta)(s\mat{E} - \mat{A}(\vartheta))^{-1} \mat{B}
\end{equation}
matches~\eqref{eq:gsk_structure}. We leave this tedious verification to the Reader, see also~\cite{Triverio2009}.

%%%%%%%%%%%%%%%%%%%%%%%%%%%%%%%%%%%%%%%%%%%%%%%%%%%%%%%%%%%
\section{Checking Passivity of Parameterized Models}\label{sec:check}
%%%%%%%%%%%%%%%%%%%%%%%%%%%%%%%%%%%%%%%%%%%%%%%%%%%%%%%%%%%

Based on the descriptor form~\eqref{eq:descriptor} of the model transfer function, we can formulate an algorithm for checking its uniform passivity, with the objective of a precise localization of eventual passivity violations.

%%%%%%%%%%%%%%%%%%%%%%%%%%%%%%%%%%%%%%%%%%%%%%%%%%%%%%%%%%%
\subsection{Passivity Check of Univariate Models}\label{sec:check_univariate}
%%%%%%%%%%%%%%%%%%%%%%%%%%%%%%%%%%%%%%%%%%%%%%%%%%%%%%%%%%%

Let us consider the univariate model obtained from~\eqref{eq:descr_tf} by ``freezing'' the parameter $\vartheta$. Checking the passivity of such a model is a standard problem~\cite{PM_book}. The most effective tool to perform this check is the so-called Hamiltonian matrix associated to the model~\cite{Grivet04pass,BBK89,LMI}, which in the case of our descriptor form becomes a Skew-Hamiltonian/Hamiltonian (SHH) matrix pencil, also denoted Generalized Hamiltonian Pencil~\cite{NWCKC2008,ZZNW2010a,ZZNW2010b,ZZNW2010c}. We define the two block-matrices
\begin{equation}\label{eq:ham_pencil}
\mat{M}(\vartheta) = \begin{bmatrix}
	\mat{A}(\vartheta) &  \mat{B} \mat{B}^\tran \\
	- \mat{C}^\tran(\vartheta) \mat{C}(\vartheta) & -\mat{A}^\tran(\vartheta)
\end{bmatrix}, \quad 
\mat{K} = \begin{bmatrix}
	\mat{E} &  \mat{0} \\
	\mat{0} & \mat{E}^\tran
\end{bmatrix}
\end{equation}
where $\mat{M}(\vartheta)$ has Hamiltonian structure and $\mat{K}$ is skew-Hamiltonian, forming the SHH matrix pencil $(\mat{M}(\vartheta),\mat{K})$. We have the following result:

\noindent
{\bf Theorem 1.} {\em Assume that the pencil $(\mat{A}(\vartheta),\mat{E})$ has no purely imaginary eigenvalues. Then, the following conditions are equivalent:
\begin{itemize}
	\item $\sigma=1$ is a singular value of $\model{\mat{H}}(\jj\omega_0;\vartheta)$ with $|\omega_0|<+\infty$;
	\item $\jj\omega_0$ is an eigenvalue of the pencil $(\mat{M}(\vartheta),\mat{K})$, i.e., there exist some vector $\vet{v}_0 \neq \vet{0}$ such that
	$$
		\mat{M}(\vartheta) \vet{v}_0 = \jj\omega_0 \mat{K} \vet{v}_0
	$$
\end{itemize}
}
\noindent
The proof is straightforward and follows the same flow as the proof of Theorem 1 in~\cite{BBK89}. See also~\cite{PM_book,ZZNW2010a,ZZNW2010c}.

Matrix $\mat{K}$ is singular, since $\mat{E}$ is singular, with ${\rm dim}\,{\rm ker}(\mat{E}) = P$ and ${\rm dim}\,{\rm ker}(\mat{K}) = 2P$. Therefore, we expect that the eigenspectrum of the pencil $(\mat{M}(\vartheta),\mat{K})$ includes at least $2P$ infinite eigenvalues. We recall that infinite eigenvalues are characterized as pairs $(\alpha,\beta)$ such that
\begin{equation}\label{eq:eig_pencil}
	\alpha \mat{M}(\vartheta) \vet{v} = \beta \mat{K} \vet{v}
\end{equation}
for some vector $\vet{v}$, with $\alpha = 0$. These eigenvalues will be disregarded in the following, and we will consider only the finite eigenvalues of the pencil as
\begin{equation}
	\lambda = \beta/\alpha,\quad \alpha\neq 0
\end{equation}
that satisfy~\eqref{eq:eig_pencil} for some $\vet{v}\neq \vet{0}$. Of course, all these eigenvalues depend on the considered parameter value $\vartheta$. Therefore, the (finite part of the) SHH eigenspectrum is parameter-dependent
\begin{equation}
	\Lambda(\vartheta) = \{\lambda(\vartheta) = \beta(\vartheta)/\alpha(\vartheta): \alpha(\vartheta)\neq 0\}.
\end{equation}
We remark that, due to the SHH structure of the pencil, the set $\Lambda(\vartheta)$ for each $\vartheta$ is symmetric with respect to both real and imaginary axis, so that eigenvalues are either occurring in pairs $\{\lambda,-\lambda\}$ if they are purely real or purely imaginary, or quadruples $\{\lambda,-\lambda,\lambda^*,-\lambda^*\}$ if they are complex-valued with a nonvanishing real and imaginary part~\cite{SHHlarge,Watkins2004,benner2002}.

\begin{figure}
	\centerline{\includegraphics[width=\columnwidth]{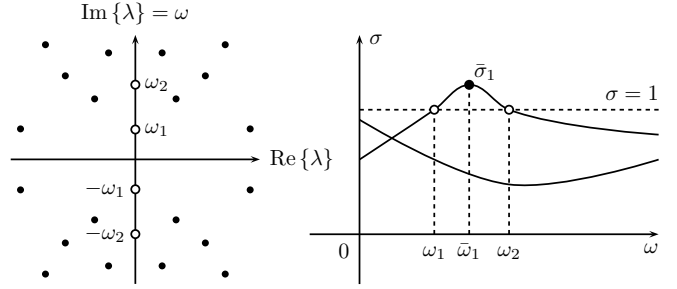}}
	\caption{Graphical illustration of Theorem~1. Left panel: finite SHH eigenvalues (empty dots highlight purely imaginary eigs); right panel: singular value trajectories, which cross the unit threshold at frequencies $\omega_i$ corresponding to purely imaginary SHH eigenvalues. Local singular value maxima $\bar{\sigma}_i$ occur at frequencies $\bar{\omega}_i$. For this example, $P=2$, $\nu=2$, $\mathcal{I}_p=\{0,2\}$ and $\mathcal{I}_{np}=\{1\}$.}
	\label{fig:univariate_check}
\end{figure}

Based on Theorem~1, if $\Lambda(\vartheta)$ includes a purely imaginary eigenvalue $\jj\omega_0$, then the transfer matrix of the model has a singular value $\sigma(\jj\omega_0,\vartheta)=1$. We conclude that the singular value trajectory for fixed $\vartheta$ in a neighborhood of $\jj\omega_0$ crosses the unit threshold, which is the critical condition leading to a passivity violation, see~\eqref{eq:BR_cond}. This violation will be for $\omega>\omega_0$ if $\sigma$ increases with frequency, and vice versa. Figure~\ref{fig:univariate_check} illustrates the relationship between imaginary SHH eigenvalues and localized passivity violations based on singular value trajectories.

As proposed in~\cite{Grivet04pass}, we define the two sets
\begin{equation}
	\chi(\vartheta) =\{\omega_i(\vartheta)\}_{i=1}^{\nu(\vartheta)}, \quad \bar{\chi}(\vartheta) =  \{0\} \cup \chi(\vartheta) \cup \{+\infty \}
\end{equation}
collecting the frequencies $\omega_i(\vartheta)>0$ corresponding to the the $\nu(\vartheta)$ purely imaginary SHH eigenvalues $\lambda_i(\vartheta) = \jj\omega_i(\vartheta)$ with positive (finite) imaginary part, sorted in ascending order. The augmented set $\bar{\chi}(\vartheta)$ includes also the DC point $\omega_0=0$ and the infinte frequency $\omega=+\infty$. The elements of $\bar{\chi}(\vartheta)$ thus induce a subdivision of the positive frequency axis $[0,+\infty)$ into $\nu(\vartheta)+1$ disjoint subbands
\begin{equation}
	\Omega_i(\vartheta) = (\omega_i(\vartheta),\omega_{i+1}(\vartheta)),\quad i=0,\dots,\nu(\vartheta),
\end{equation}
where we defined $\omega_0(\vartheta)=0$ and $\omega_{\nu(\vartheta)+1}(\vartheta) = +\infty$. Since all intersections of some singular value trajectory with the threshold $\sigma=1$ is captured in the set $\chi(\vartheta)$, each subband $\Omega_i(\vartheta)$ can be flagged either as {\em locally passive} (if all singular values are less than 1 in this band) or {\em locally not passive} (otherwise), by splitting the corresponding index sets into $i\in\mathcal{I}_{p}(\vartheta)$ and $i\in\mathcal{I}_{np}(\vartheta)$, respectively. The worst-case passivity violation for each non-passive band is defined as the maximum singular value $\bar{\sigma}_{i}(\vartheta)$, occurring at some frequency $\bar{\omega}_i(\vartheta) \in \Omega_i(\vartheta)$ with $i\in\mathcal{I}_{np}(\vartheta)$. Numerical estimates of such local maxima $(\bar{\omega}_i(\vartheta),\bar{\sigma}_{i}(\vartheta))$ are easily determined through local sampling. See Figure~\ref{fig:univariate_check} for a graphical illustration.

%%%%%%%%%%%%%%%%%%%%%%%%%%%%%%%%%%%%%%%%%%%%%%%%%%%%%%%%%%%
\subsection{Uniform Passivity Check of Parameterized Models}\label{sec:check_multivariate}
%%%%%%%%%%%%%%%%%%%%%%%%%%%%%%%%%%%%%%%%%%%%%%%%%%%%%%%%%%%

The procedure discussed in Section~\ref{sec:check_univariate} allows a passivity check of univariate models throughout the frequency axis, without any need of sampling the singular values of the transfer function, but requiring only an algebraic determination of some SHH eigenvalues. In this Section, we enrich this test by extending its scope to a uniform passivity check throughout the parameter range $\vartheta\in\Theta$. The main tool that we consider is the auxiliary function $\psi(\vartheta)$, defined as
\begin{equation}
	\psi(\vartheta) = \min_{\lambda(\vartheta)\in \Lambda(\vartheta)}   \frac{|\,\Re{\lambda(\vartheta)}|}{\rho(\vartheta)} 
\end{equation}
where $\rho(\vartheta) = \max_{\lambda(\vartheta)\in \Lambda(\vartheta)} |\lambda(\vartheta)|$ is the spectral radius of the SHH pencil~\eqref{eq:ham_pencil}, computed considering only the finite eigenvalues. This function vanishes when the set $\chi(\vartheta)$ is non-empty, denoting the presence of purely imaginary eigenvalues. Conversely, if $\chi(\vartheta)$ is empty and there are no imaginary eigenvalues, so that the univariate model is passive for the considered value of $\vartheta$, the function $\psi(\vartheta)$ measures the distance of the SHH eigenspectrum from the imaginary axis, normalized to the largest eigenvalue magnitude. This normalization makes the test less sensitive to the finite numerical precision in the eigenvalue computation, as well as independent on the units and/or normalizations used for the frequency variable.

\begin{figure}
\centerline{\includegraphics[width=\columnwidth]{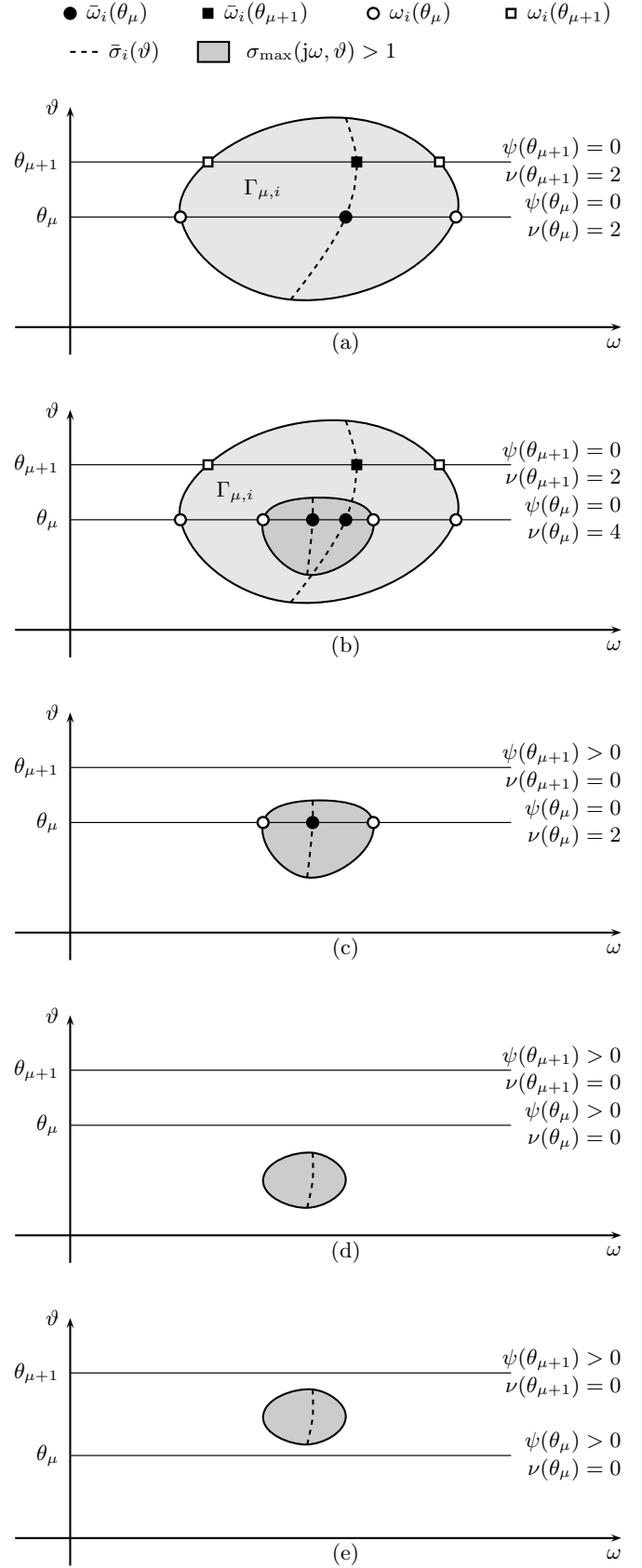}}
\caption{Adaptive parameter sampling, see main text for details.}
\label{fig:sampling_theta}
\end{figure}

The proposed check performs an adaptive sampling of $\psi(\vartheta)$ within the range $\vartheta \in \Theta = [\vartheta_{\min},\vartheta_{\max}]$. We start with an initial uniform partition into $\bar{\mu}$ subintervals with endpoints
\begin{equation}\label{eq:initial_sampling}
	\theta_\mu = \vartheta_{\min} + \frac{\mu}{\bar{\mu}} (\vartheta_{\max}-\vartheta_{\min}),\quad \mu=0,\dots,\bar{\mu},
\end{equation}
and we apply the univariate passivity check by computing the SHH eigenvalues at each $\theta_\mu$, as discussed in Section~\ref{sec:check_univariate}. An adaptive refinement process is then constructed, based on the following observations. Considering a single subinterval $\Theta_\mu = [\theta_\mu,\theta_{\mu+1}]$,
\begin{itemize}
	\item if $\psi(\vartheta)=0,\;\forall \vartheta\in \Theta_\mu$, the model is not passive in this subinterval due to the presence if imaginary SHH eigenvalues. For each $\vartheta$, such eigenvalues delimit at least one frequency band $\Omega_i(\vartheta)$ where at least one singular value of the model $\model{\mat{H}}(\jj\omega;\vartheta)$ exceeds one, with a local maximum $\bar{\sigma}_i(\vartheta)$ occurring at frequency $\bar{\omega}_i(\vartheta)$.
	\item if $\psi(\vartheta)>0,\;\forall \vartheta\in \Theta_\mu$, the model is uniformly passive in $\Theta_\mu$.
\end{itemize}
Assuming now that only information at the edges $\{\theta_\mu,\theta_{\mu+1}\}$ of $\Theta_\mu$ is available, we have the following subcases (Fig.~\ref{fig:sampling_theta}):
\begin{itemize}
	\item if $\psi(\theta_\mu) = \psi(\theta_{\mu+1}) = 0$ and $\nu(\theta_\mu) = \nu(\theta_{\mu+1})$, we can infer that there exist at least one region $\Gamma_{\mu,i}$ of the frequency-parameter plane $\Gamma \subseteq [0,+\infty) \times \Theta_\mu$ where the model is uniformly non-passive (see Figure~\ref{fig:sampling_theta}a). For each region $\Gamma_{\mu,i}$, we know the local singular value maxima $\bar{\sigma}_i(\theta_\mu)$, $\bar{\sigma}_i(\theta_{\mu+1})$, and the corresponding localization frequencies $\bar{\omega}_i(\theta_\mu)$, $\bar{\omega}_i(\theta_{\mu+1})$. In addition, we know all the edges of $\Gamma_{\mu,i}$ for $\theta_\mu$ and $\theta_{\mu+1}$ from the sets $\chi(\theta_\mu)$, $\chi(\theta_{\mu+1})$. Since local passivity violations have been identified, there is no need to refine subinterval $\Theta_\mu$.	
	\item if $\psi(\theta_\mu) = \psi(\theta_{\mu+1}) = 0$ and $\nu(\theta_\mu) \neq \nu(\theta_{\mu+1})$, the number of frequency bands possibly changes when sweeping $\vartheta \in \Theta_\mu$ (see Figure~\ref{fig:sampling_theta}b). In order to obtain a precise characterization, we refine $\Theta_\mu$ by adding the new point
	\begin{equation}\label{eq:new_point}
	\theta_{\mu+\onehalf} = \onehalf(\theta_\mu+\theta_{\mu+1}).
	\end{equation}
	\item if $\psi(\theta_\mu) = 0$ and $\psi(\theta_{\mu+1}) > 0$, or conversely $\psi(\theta_\mu) > 0$ and $\psi(\theta_{\mu+1}) = 0$, we have a transition from a passive to a non-passive model while sweeping $\vartheta \in \Theta_\mu$ (see Figure~\ref{fig:sampling_theta}c). Therefore, we need to refine $\Theta_\mu$ through~\eqref{eq:new_point} in order to track the particular $\vartheta_*$ where the onset of a passivity violation occurs.
	\item if both $\psi(\theta_\mu) > 0$ and $\psi(\theta_{\mu+1}) > 0$, we can have two subcases:
	\begin{itemize}
		\item $\psi(\vartheta) > 0$ throughout $\Theta_\mu$: model is uniformly passive in $\Theta_\mu$ and no refinement is necessary (Figure~\ref{fig:sampling_theta}d);
		\item $\psi(\vartheta)$ vanishes in some subinterval $\Theta_*\subseteq \Theta_\mu$, denoting a passivity violation that is not visible from the edges of $\Theta_\mu$ (Figure~\ref{fig:sampling_theta}e). We need to refine $\Theta_\mu$.
	\end{itemize}
	These two cases are here discriminated by computing $\psi(\theta_{\mu+\onehalf})$, and by constructing the first-order interpolation error
	\begin{equation}
		\varepsilon_{\mu+\onehalf} = \left|\psi(\theta_{\mu+\onehalf}) - \onehalf[\psi(\theta_{\mu})+\psi(\theta_{\mu+1})]\right|\,.
	\end{equation}
	Refinement is applied if
	\begin{equation}
		\varepsilon_{\mu+\onehalf} > \gamma \left|  \psi(\theta_{\mu+\onehalf}) \right|,
	\end{equation}
	equivalently, when the estimate of the midpoint through linear interpolation is not sufficient to infer that $\psi(\vartheta)$ is uniformly positive within $\Theta_\mu$. The parameter $\gamma$ can be used to tune the selectivity of the refinement test, here we use the conservative value $\gamma=0.2$.
\end{itemize}
We perform a total number $M$ of refinement passes. At each pass, the new points $\theta_{\mu+\onehalf}$ are added to the previous subset $\{ \theta_\mu,\;\mu=0,\dots,\bar{\mu}\}$, which is resorted in ascending order and reindexed, by suitably redefining $\bar{\mu}$. Algorithm~\ref{al:pass_check} summarizes the proposed multivariate passivity check in pseudocode form (step~10 embeds the determination of new parameter samples $\theta_{\mu}^m$ at each refinement pass $m$, according to the above adaptive process). Throughout this work, we use $M=10$.

A final remark is in order about the initial sampling~\eqref{eq:initial_sampling}. Here, we need to make sure that we do not miss important information due to an excessively coarse sampling. On the other hand, we do not want to spend unnecesary computations if the number of samples is too large. The number of initial subbands $\bar{\mu}$ depends in fact on the model variations throughout the parameter space, which in turn is directly related to the type and number of adopted basis functions $\xi_\ell(\vartheta)$ for model parameterization in~\eqref{eq:param_coeff}. In this work, we assume entire-domain smooth basis functions (polynomials, orthogonal polynomials, or trigonometric polynomials). Due to the smooth parameterization, it is sufficient to consider the heuristic rule $\bar{\mu}=\kappa\, \bar{\ell}$, with $\kappa>1$. For all documented examples, we set $\kappa=4$. 

\begin{algorithm}
\caption{Passivity check of parameterized models}
\label{al:pass_check}
\begin{algorithmic}[1]
	\Require frequency basis $\varphi_n$ for $n=0,\dots,\bar{n}$;
	\Require parameter basis $\xi_\ell$ for $\ell=1,\dots,\bar{\ell}$;
	\Require model coefficients $\mat{R}_{n,\ell}$, $r_{n,\ell}$  in~\eqref{eq:gsk_structure} and~\eqref{eq:param_coeff};
	\Require control parameters $\vartheta_{\min}$, $\vartheta_{\max}$, $\gamma$, $\kappa$, $M$;
	\State set $m=0$ and number of initial samples $\bar{\mu}_0=\kappa\, \bar{\ell}$;
	\State set initial samples $\mathcal{S}_0 = \{\theta_{\mu}^0,\,\mu=0,\dots,\bar{\mu}_0\}$ as in~\eqref{eq:initial_sampling};
	\Repeat
		\For{$\mu=1,\dots,\bar{\mu}_m$}
			\State construct SHH pencil $(\mat{M}(\theta_{\mu}^m),\mat{K})$;
			\State find imaginary SHH eigenvalues $\omega_i(\theta_{\mu}^m)$;
			\State extract local singular value maxima $(\bar{\omega}_{\mu,i}^m,\bar{\sigma}_{\mu,i}^m)$;
		\EndFor
		\State $m \leftarrow m+1$
		\State determine new samples $\mathcal{S}_m = \{\theta_{\mu}^m,\,\mu=1,\dots,\bar{\mu}_m\}$;
	\Until{$\mathcal{S}_m=\emptyset$ or $m=M$}
	\State \textbf{return} passivity violations $\mathcal{V} = \cup_m \{ (\bar{\omega}_{\mu,i}^m,\bar{\sigma}_{\mu,i}^m),\,\forall \mu,i \}$.
\end{algorithmic}
\end{algorithm}

\begin{figure}
	\centerline{\includegraphics[width=\columnwidth,clip]{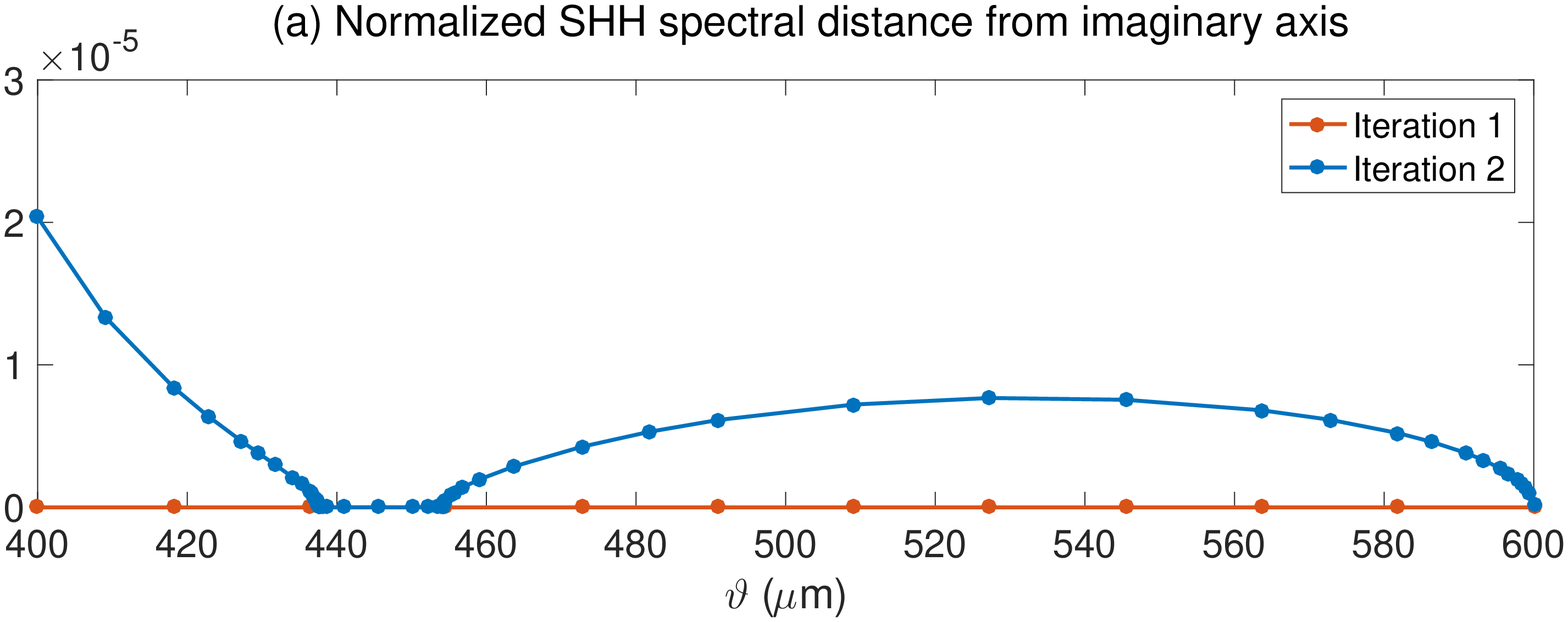}}
	\centerline{\includegraphics[width=\columnwidth,clip]{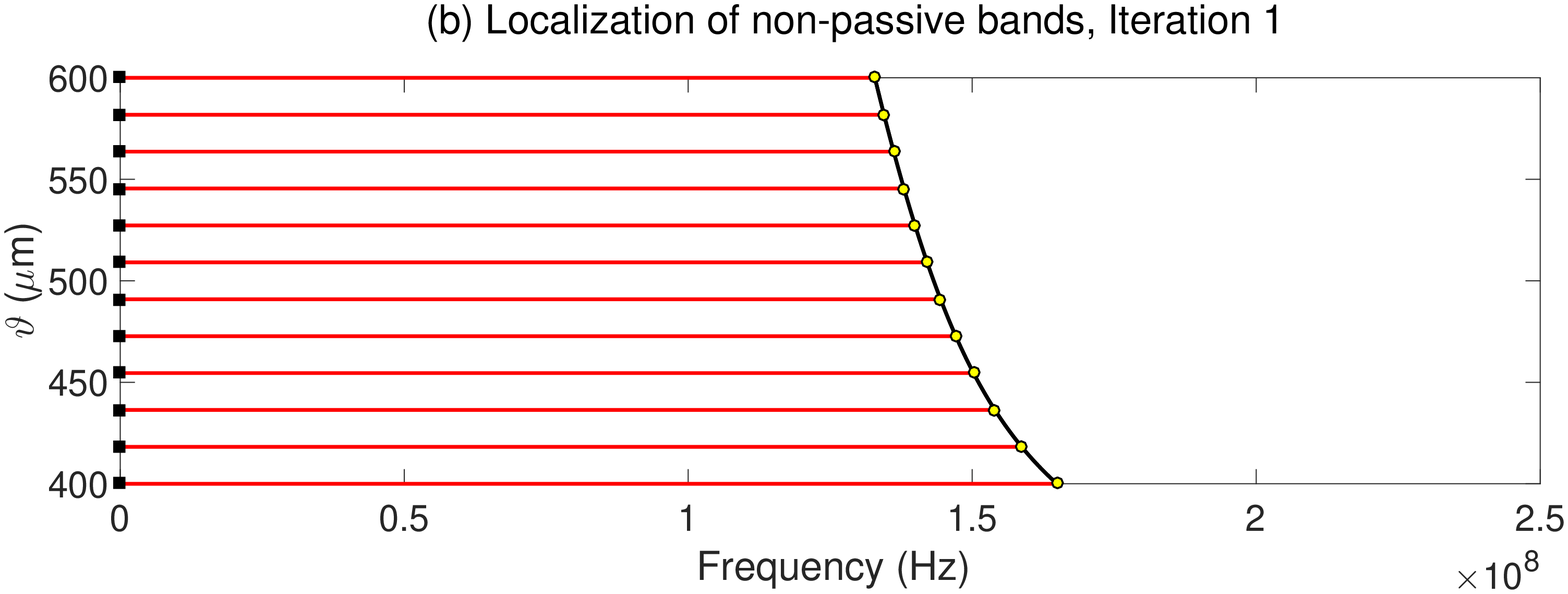}}
	\centerline{\includegraphics[width=\columnwidth,clip]{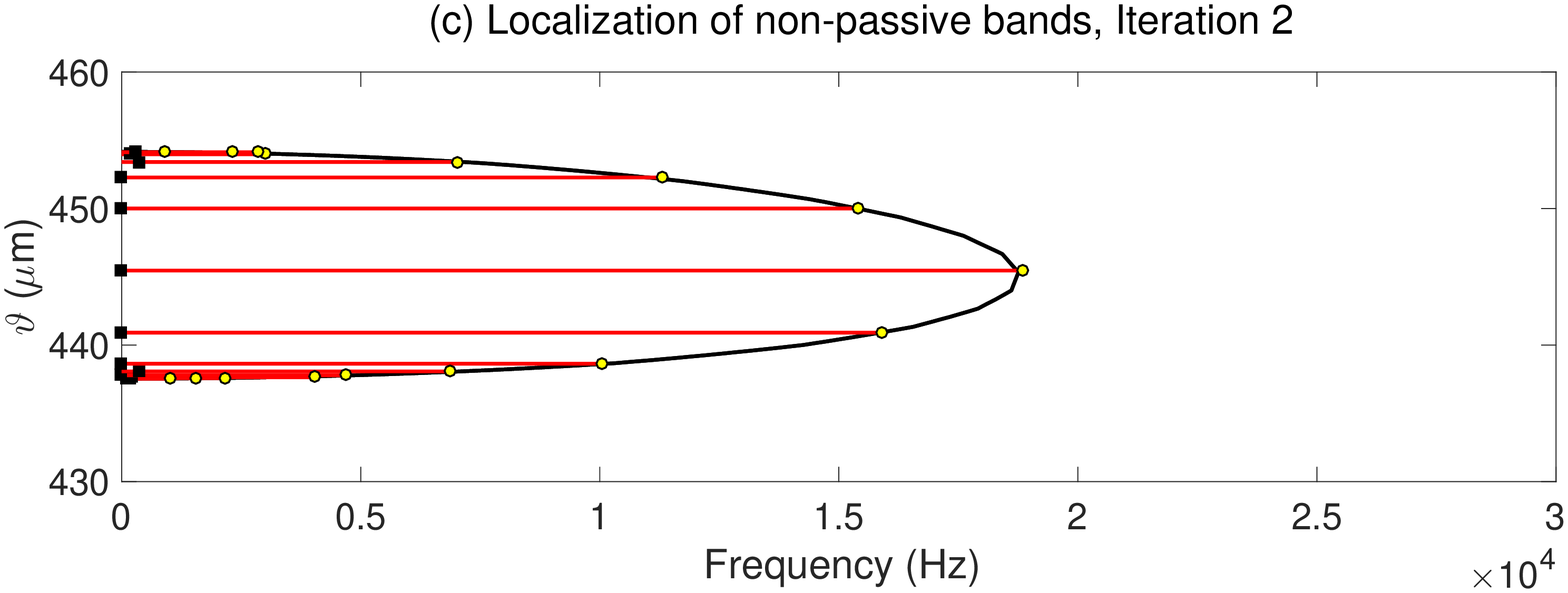}}
	\centerline{\includegraphics[width=\columnwidth,clip]{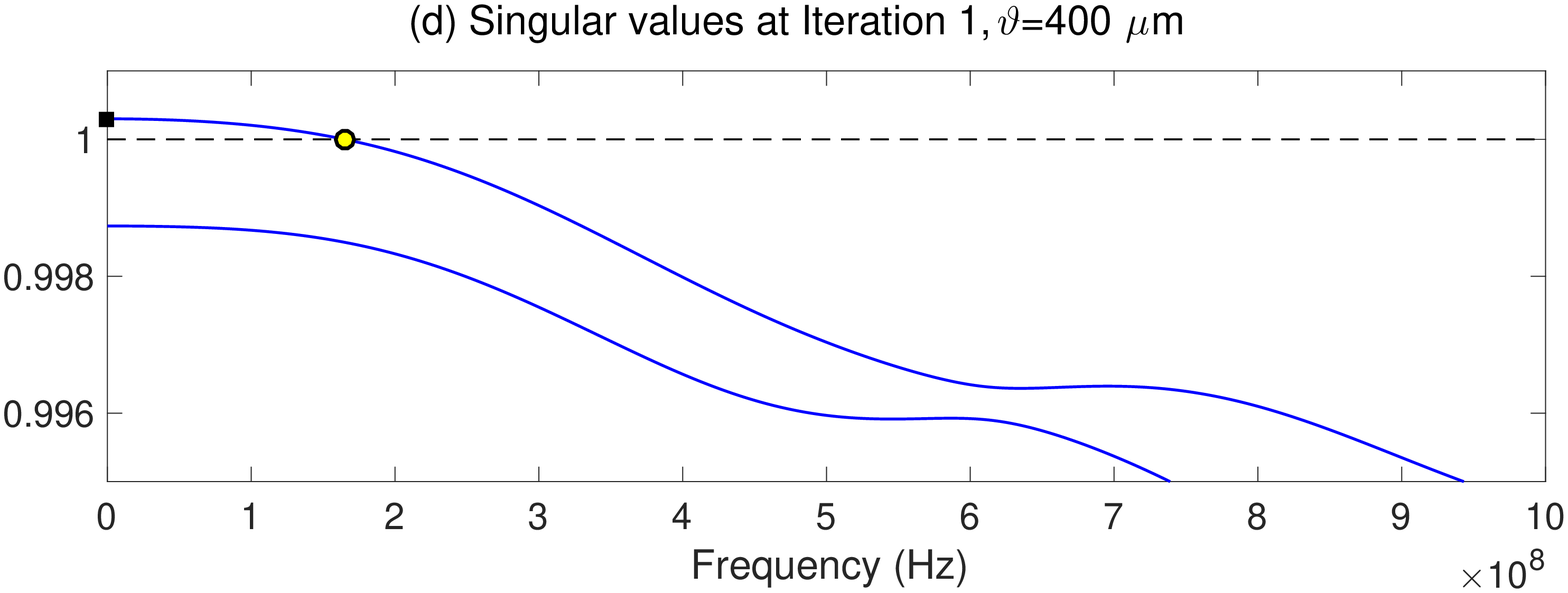}}
	\caption{Checking uniform passivity of two PCB link macromodels parameterized by the via antipad radius $\vartheta=r$, see Section~\protect\ref{sec:Slink}). The two models correspond to two different iterations of the passivity enforcement loop, discussed in Section~\protect\ref{sec:enforcement}. Panel (a): adaptively computed samples of $\psi(\theta_\mu)$ for the two models. Panels (b) and (c): for all computed $\theta_\mu$, each non-passive frequency band is highligthed with a red line; the yellow dots highlight the frequencies of imaginary SHH eigenvalues; the black squares are the local singular value maxima; the solid black line is an approximation of the contour line for $\sigma=1$ of the singular value trajectories $\sigma(\jj\omega;\vartheta)$ (not required by the algorithm and depicted only for illustration purpose). Panel (d): singular value plot of model at Iteration~1.}
	\label{fig:Slink_pass_check}
\end{figure}

Figure~\ref{fig:Slink_pass_check} illustrates the proposed passivity check on a practical example (discussed in more detail in Section~\ref{sec:Slink}). Two different models of a PCB interconnect link are processed by the proposed passivity check algorithm. Figure~\ref{fig:Slink_pass_check}(a) plots the function $\psi(\vartheta)$ for the two models. The model no.~1 is not passive for any value of $\vartheta$, hence $\psi(\theta_\mu)=0$ for all computed $\mu$ (the red dots). Correspondingly, Figure~\ref{fig:Slink_pass_check}(b) shows that for all $\theta_\mu$ samples, there is at least one imaginary SHH eigenvalue (yellow dots) that defines a frequency band where at least one singular value of the model is larger than one (red lines). Local singular value maxima are close to DC (black squares). Note that, since the number of detected imaginary SHH eigenvalues is constant for all $\theta_\mu$, no adaptive refinement is necessary. The blue curve in Figure~\ref{fig:Slink_pass_check}(a) shows instead $\psi(\vartheta)$ for model no.~2, which is locally not passive only in a restricted range of $\vartheta\in\Theta_*\approx\, [437,454]~\mu$m. Correspondingly, imaginary SHH eigenvalues are only found in this range: see Figure~\ref{fig:Slink_pass_check}(c), where adaptive refinement is clearly visible near the transitions between $\psi(\vartheta)=0$ and $\psi(\vartheta)>0$, which define the endpoints of $\Theta_*$. Finally, Figure~\ref{fig:Slink_pass_check}(d) shows the frequency-dependent singular values of model no.~1 for $\vartheta=400~\mu$m, where the frequency of the unique imaginary SHH eigenvalue is highlighted by a yellow dot.

%%%%%%%%%%%%%%%%%%%%%%%%%%%%%%%%%%%%%%%%%%%%%%%%%%%%%%%%%%%
\section{Enforcing Passivity of Parameterized Models}\label{sec:enforcement}
%%%%%%%%%%%%%%%%%%%%%%%%%%%%%%%%%%%%%%%%%%%%%%%%%%%%%%%%%%%

Based on the multivariate passivity characterization discussed in Section~\ref{sec:check_multivariate}, we are now ready to formulate our proposed passivity enforcement algorithm. A perturbed model based on~\eqref{eq:pert_coefficients} is defined as
\begin{equation}\label{eq:pert_model}
	\pert{\model{\mat{H}}}(s;\vartheta) = \model{\mat{H}}(s;\vartheta) + \Delta\model{\mat{H}}(s;\vartheta)
\end{equation}
with
\begin{equation}\label{eq:model_perturbation}
	\Delta\model{\mat{H}}(s;\vartheta)  = \dfrac{\Delta\textsf{\textbf{N}}(s,\vartheta)}{\mathsf{D}(s,\vartheta)} 
		 = \dfrac{\sum_{n=0}^{\bar{n}} \sum_{\ell=1}^{\bar{\ell}} \Delta\mat{R}_{n,\ell}\,\xi_\ell(\vartheta)\, \varphi_n(s)}%
			{\sum_{n=0}^{\bar{n}}  \sum_{\ell=1}^{\bar{\ell}} r_{n,\ell}\,\xi_\ell(\vartheta)\, \varphi_n(s)}
\end{equation}
The coefficient perturbations $\Delta\mat{R}_{n,\ell}$ will be computed such that the singular values of the perturbed model that are larger than one will be displaced below the unit threshold. Such coefficients will be collected in a global vector of decision varibles
\begin{equation}\label{eq:decision_variables}
	\xx = \begin{bmatrix} \xx_{1,1}^\tran, \cdots, \xx_{i,j}^\tran, \cdots, \xx_{P,P}^\tran  \end{bmatrix}^\tran,
\end{equation}
where
\begin{equation}\label{eq:decision_variables_ij}
	\xx_{i,j} = \begin{bmatrix}  (\Delta\mat{R}_{0,1})_{i,j}, \cdots, (\Delta\mat{R}_{n,\ell})_{i,j}, \cdots, (\Delta\mat{R}_{\bar{n},\bar{\ell}})_{i,j}  \end{bmatrix}^\tran
\end{equation}
collects the $(i,j)$ entries of all matrices $\Delta\mat{R}_{n,\ell}$ with a suitable ordering. We have $\xx_{i,j} \in \mathbb{R}^{(\bar{n}+1)\bar{\ell}}$ and $\xx\in\mathbb{R}^Q$, with $Q=P^2 (\bar{n}+1)\bar{\ell}$.

%%%%%%%%%%%%%%%%%%%%%%%%%%%%%%%%%%%%%%%%%%%%%%%%%%%%%%%%%%%
\subsection{Building Algebraic Passivity Constraints}\label{sec:constraints}
%%%%%%%%%%%%%%%%%%%%%%%%%%%%%%%%%%%%%%%%%%%%%%%%%%%%%%%%%%%

Let us consider a single local singular value maximum $\bar{\sigma}_{\mu,i} = \bar{\sigma}_{i}(\theta_\mu) > 1$ occurring at frequency $\bar{\omega}_{\mu,i} = \bar{\omega}_i(\theta_\mu)$, as resulting from the passivity check of Section~\ref{sec:check_multivariate}. Model perturbation leads to a perturbation of this singular value, which under first-order approximation reads~\cite{bai}
\begin{equation}
	\pert{\sigma}_{\mu,i} \approx \bar{\sigma}_{\mu,i} + \Re{ \vet{u}_{\mu,i}^\herm \, \Delta\model{\mat{H}}(\jj\bar{\omega}_{\mu,i};\theta_\mu) \, \vet{v}_{\mu,i} },
\end{equation}
where $\vet{u}_{\mu,i},\vet{v}_{\mu,i}$ are the left and right singular vectors of $\model{\mat{H}}(\jj\bar{\omega}_{\mu,i};\theta_\mu)$ associated to $\bar{\sigma}_{\mu,i}$. Forcing this singular value to be less than one leads to the inequality constraint
\begin{equation}\label{eq:pass_constraint_base}
	 \Re{ \vet{u}_{\mu,i}^\herm \, \Delta\model{\mat{H}}(\jj\bar{\omega}_{\mu,i};\theta_\mu) \, \vet{v}_{\mu,i} } \leq 1 - \bar{\sigma}_{\mu,i}.
\end{equation}
Defining now
\begin{equation}
	\vet{a}_{\mu,i} =  \begin{bmatrix}  a_{\mu,i;0,1}, \cdots, a_{\mu,i;n,\ell}, \cdots, a_{\mu,i;\bar{n},\bar{\ell}}  \end{bmatrix}^\tran
\end{equation}
with the same ordering as in~\eqref{eq:decision_variables_ij}, where
\begin{equation}
	a_{\mu,i;n,\ell} = \frac{\xi_\ell(\theta_\mu)\, \varphi_n(\jj\bar{\omega}_{\mu,i})}{\mathsf{D}(\jj\bar{\omega}_{\mu,i},\theta_\mu)},
\end{equation}
we can cast~\eqref{eq:pass_constraint_base} in algebraic form
\begin{equation}\label{eq:pass_constraint_single}
	\vet{p}_{\mu,i}^\tran \, \xx \leq 1 - \bar{\sigma}_{\mu,i},
\end{equation}
where
\begin{equation}
	\vet{p}_{\mu,i}^\tran = \Re{ ( \vet{v}_{\mu,i}^\tran \otimes \vet{u}_{\mu,i}^\herm ) \otimes \vet{a}_{\mu,i}^\tran }.
\end{equation}

The constraint~\eqref{eq:pass_constraint_base} operates on an individual singular value maximum $\bar{\sigma}_{\mu,i}$, attempting to reduce its value to be less than one (since based on a first-order singular value perturbation, this enforcement is not exact but only approximate). If multiple singular value maxima are perturbed concurrently, it is sufficient to add a constraint~\eqref{eq:pass_constraint_base} for each $\mu,i$.

%%%%%%%%%%%%%%%%%%%%%%%%%%%%%%%%%%%%%%%%%%%%%%%%%%%%%%%%%%%
\subsection{Preserving Model Accuracy}\label{sec:cost}
%%%%%%%%%%%%%%%%%%%%%%%%%%%%%%%%%%%%%%%%%%%%%%%%%%%%%%%%%%%

Enforcing constraint~\eqref{eq:pass_constraint_base} does not guarantee that the perturbed model remains accurate. Hence the need of constructing a cost function that casts in algebraic form a suitable norm of the model perturbation. We define such cost function as
\begin{equation}\label{eq:cost_function_def}
	\mathcal{E}^2 = \sum_{i,j=1}^P \mathcal{E}_{i,j}^2,
\end{equation}
where
\begin{equation}\label{eq:cost_function_def_ij}
	\mathcal{E}_{i,j}^2 = \sum_{k=1}^{\bar{k}} \sum_{m=1}^{\bar{m}} w_{i,j;k,m}^2 \abs{\Delta\model{H}_{i,j}(\jj2\pi f_k,\vartheta_m) }^2
\end{equation}
denotes the $(i,j)$-th entry of the squared model perturbation at frequency $f_k$ and parameter $\vartheta_m$, weighted by the possibly frequency-, parameter-, and entry-dependent weight $w_{i,j;k,m}$.
Defining now
\begin{equation}
	b_{k,m;n,\ell} = \frac{\xi_\ell(\vartheta_m)\, \varphi_n(\jj2\pi f_k)}{\mathsf{D}(\jj2\pi f_k,\vartheta_m)}
\end{equation}
and
\begin{equation}
	\vet{b}_{k,m}^\tran =
	\begin{bmatrix}
		b_{k,m;0,1},\cdots,b_{k,m;n,\ell},\cdots,b_{k,m;\bar{n},\bar{\ell}}
	\end{bmatrix}
\end{equation}
we can cast the elementwise cost function~\eqref{eq:cost_function_def_ij} as
\begin{equation}\label{eq:cost_function_ij}
	\mathcal{E}_{i,j}^2 = \norm{ \mat{F}_{i,j}\, \xx_{i,j} }_2^2,
\end{equation}
where
\begin{equation}
	\mat{F}_{i,j} = \begin{bmatrix}
		\Re{\widetilde{\mat{F}}_{i,j}} \\
		\Im{\widetilde{\mat{F}}_{i,j}}
	\end{bmatrix},\quad
	\widetilde{\mat{F}}_{i,j} = \begin{bmatrix}
		w_{i,j;1,1}\,\vet{b}_{1,1}^\tran \\
		\vdots \\
		w_{i,j;k,m}\,\vet{b}_{k,m}^\tran \\
		\vdots \\
		w_{i,j;\bar{k},\bar{m}}\,\vet{b}_{\bar{k},\bar{m}}^\tran
	\end{bmatrix}
\end{equation}
Note that $\mat{F}_{i,j} \in \mathbb{R}^{2\bar{k}\bar{m} \times (\bar{n}+1)\bar{\ell}}$ collects as many rows as available frequency and parameter samples. The row size can thus be large, since usually $2\bar{k}\bar{m} \gg (\bar{n}+1)\bar{\ell}$. An equivalent compressed form of~\eqref{eq:cost_function_ij} reads
\begin{equation}\label{eq:cost_function_qr_ij}
	\mathcal{E}_{i,j}^2 = \norm{ \mat{\Psi}_{i,j}\, \xx_{i,j} }_2^2,
\end{equation}
where $\mat{\Psi}_{i,j}\in\mathbb{R}^{(\bar{n}+1)\bar{\ell} \times (\bar{n}+1)\bar{\ell}}$ is obtained through an ``economy-size'' QR factorization of $\mat{F}_{i,j} = \mat{Q}_{i,j} \mat{\Psi}_{i,j}$, where $\mat{Q}_{i,j}^\tran \mat{Q}_{i,j} = \mat{I}$. Finally~\eqref{eq:cost_function_def} can be cast as
\begin{equation}\label{eq:cost_function}
	\mathcal{E}^2 = \norm{ \mat{\Psi}\, \xx }_2^2,
\end{equation}
where
\begin{equation}\label{eq:Psi}
	\mat{\Psi} = {\rm blkdiag} \{ \mat{\Psi}_{i,j} \}_{i,j=1}^P
\end{equation}

%%%%%%%%%%%%%%%%%%%%%%%%%%%%%%%%%%%%%%%%%%%%%%%%%%%%%%%%%%%
\subsection{Iterative Passivity Enforcement}\label{sec:iterations}
%%%%%%%%%%%%%%%%%%%%%%%%%%%%%%%%%%%%%%%%%%%%%%%%%%%%%%%%%%%

A first-order singular value perturbation with minimum induced model error is achieved by minimizing~\eqref{eq:cost_function} while enforcing~\eqref{eq:pass_constraint_single}, resulting in the following constrained minimization problem
\begin{equation}\label{eq:single_iteration}
	{\min}_{\xx} \norm{ \mat{\Psi}\, \xx }_2^2 \quad \text{subject to}\quad \vet{p}_{\mu,i}^\tran \, \xx \leq 1 - \bar{\sigma}_{\mu,i},\quad \forall \mu,i.
\end{equation}
This problem is convex thus straightforward to solve, e.g., through a Primal-Dual Interior Point method~\cite{boyd2004}. Since based on a first-order approximation, the above optimization may need to be iterated until all passivity violations are removed. Algorithm~\ref{al:pass_enf} illustrates the resulting passivity enforcement scheme in pseudocode form. We see that this algorithm is a straightforward extension to the multivariate case of the singular value perturbation scheme~\cite{jnl-2008-tadvp-PassivityMethods}, which is only applicable to univariate models. The key for the proposed multivariate extension is the availability of a reliable process for detecting and extracting the parameter-dependent passivity violations of the model, as discussed in Section~\ref{sec:check}.

\begin{algorithm}
\caption{Passivity enforcement of parameterized models}
\label{al:pass_enf}
\begin{algorithmic}[1]
	\Require frequency basis $\varphi_n$ for $n=0,\dots,\bar{n}$;
	\Require parameter basis $\xi_\ell$ for $\ell=1,\dots,\bar{\ell}$;
	\Require model coefficients $\mat{R}_{n,\ell}$, $r_{n,\ell}$  in~\eqref{eq:gsk_structure} and~\eqref{eq:param_coeff};
	\Require control parameters $\vartheta_{\min}$, $\vartheta_{\max}$, $\gamma$, $\kappa$, $M$;
	\State find passivity violations $\mathcal{V}$ via Algorithm~\ref{al:pass_check};
	\While{$\mathcal{V} \neq \emptyset$}
		\State build constraint~\eqref{eq:pass_constraint_single} for each element in $\mathcal{V}$;
		\State compute matrix $\mat{\Psi}$ in~\eqref{eq:Psi};
		\State solve convex optimization problem~\eqref{eq:single_iteration};
		\State update model coefficients $\mat{R}_{n,\ell} \leftarrow \mat{R}_{n,\ell} + \Delta\mat{R}_{n,\ell}$;
		\State find passivity violations $\mathcal{V}$ via Algorithm~\ref{al:pass_check};	
	\EndWhile
	\State \textbf{return} passive model $\pert{\model{\mat{H}}}(s;\vartheta)$.
\end{algorithmic}
\end{algorithm}

%%%%%%%%%%%%%%%%%%%%%%%%%%%%%%%%%%%%%%%%%%%%%%%%%%%%%%%%%%%
\section{Examples}\label{sec:examples}
%%%%%%%%%%%%%%%%%%%%%%%%%%%%%%%%%%%%%%%%%%%%%%%%%%%%%%%%%%%

The performance of proposed passivity enforcement scheme is now demonstrated on three examples. A laptop with Intel Core i7 CPU running at 2.6~GHz with 16~GB RAM was used in all numerical simulations.

%%%%%%%%%%%%%%%%%%%%%%%%%%%%%%%%%%%%%%%%%%%%%%%%%%%%%%%%%%%
\subsection{A Printed Circuit Board Interconnect}\label{sec:Slink}
%%%%%%%%%%%%%%%%%%%%%%%%%%%%%%%%%%%%%%%%%%%%%%%%%%%%%%%%%%%

The first example we consider is a high-speed signal link (see~\cite{SchusterDesignCon2017} for a detailed description) routed on the inner layers of two Printed Circuit Boards hinged by a connector. Vertical interconnection at the feeding ports and at the connector ports is provided by four through vias. The $2\times 2$ scattering responses of the link are parameterized by the via antipad radius $\vartheta=r$ within the range $\Theta= [400,600]\,\mu$m. The raw frequency-domain scattering responses (Courtesy of Prof. Christian Schuster and Dr. Jan Preibisch, Technische Universit\"at Hamburg-Harburg, Hamburg, Germany) are obtained through a combination of a full-wave field solver (for the connector), lossy transmission-line models for the stripline segments, and a field model for the vias based on~\cite{CIM}. A total of $\bar{k}=500$ frequency samples up to 10~GHz, combined with $\bar{m}=9$ parameter samples were available for model identification.

The parameterized GSK iteration of Section~\ref{sec:notation} was applied with $\bar{n}=44$ basis poles, and using orthogonal (Chebychev) polynomials $\xi_\ell(\vartheta)$ to represent parameter variations through~\eqref{eq:param_coeff}. The number of basis functions was determined by trial and error as $\bar{\ell}=3$, corresponding to quadratic polynomials. Only one half of the parameter samples (indices $m=1,3,5,7,9$) were used to fit the model, leaving the even-numbered $m=2,4,6,8$ for a-posteriori model validation purposes. The worst-case absolute RMS error among all scattering matrix elements at both fitting and validation points resulted $8.788 \times 10^{-4}$, whereas the relative error was $3.367 \times 10^{-3}$. This model corresponds to the model at Iteration~1 depicted in Fig.~\ref{fig:Slink_pass_check}. We see from panels (a), (b), and (d) that this model is not passive, due to localized low-frequency passivity violations throughout the parameter range.

\begin{figure}
	\centerline{\includegraphics[width=\columnwidth,clip]{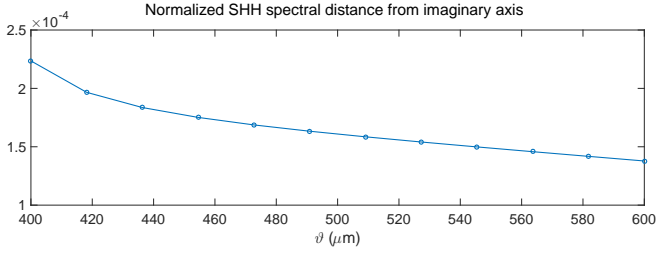}}
	\caption{Samples of $\psi(\vartheta)$ for the PCB link macromodel after passivity enforcement.}
	\label{fig:Slink_psi_passive}
\end{figure}

\begin{figure}
	\centerline{\includegraphics[width=\columnwidth,clip]{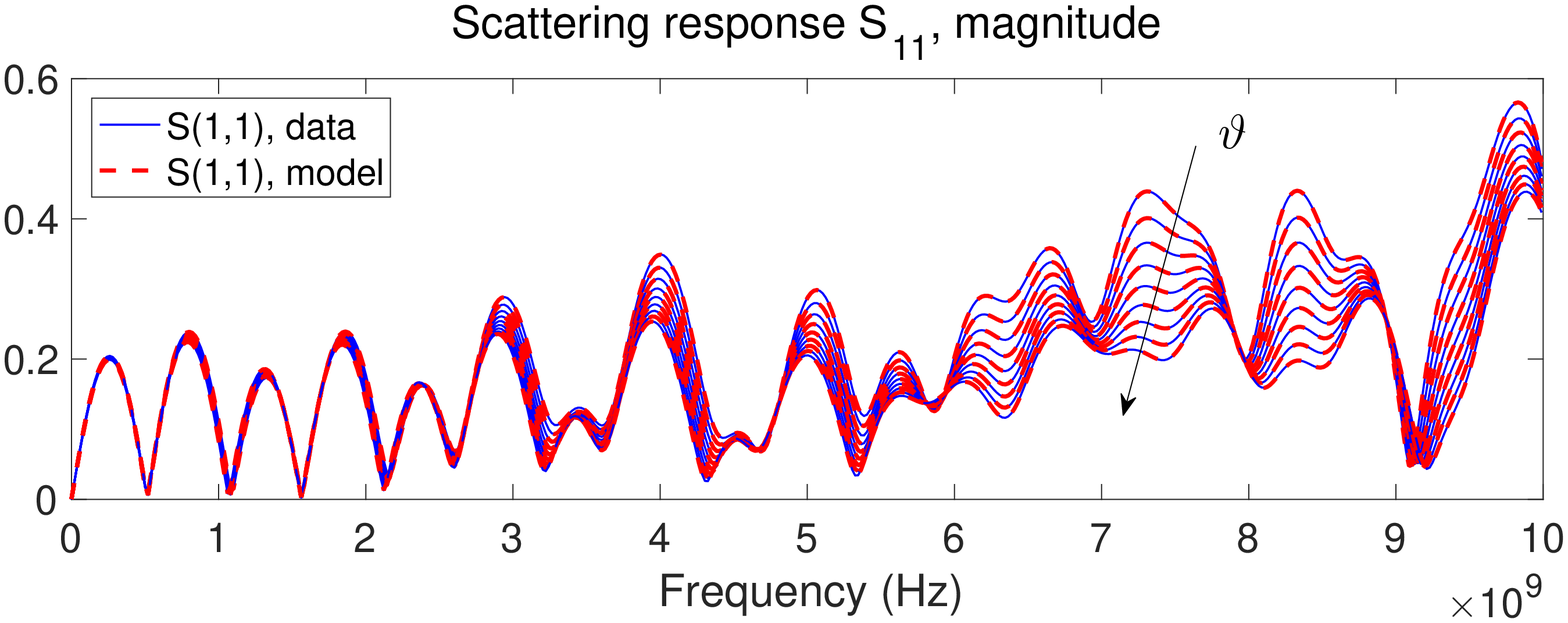}}
	\centerline{\includegraphics[width=\columnwidth,clip]{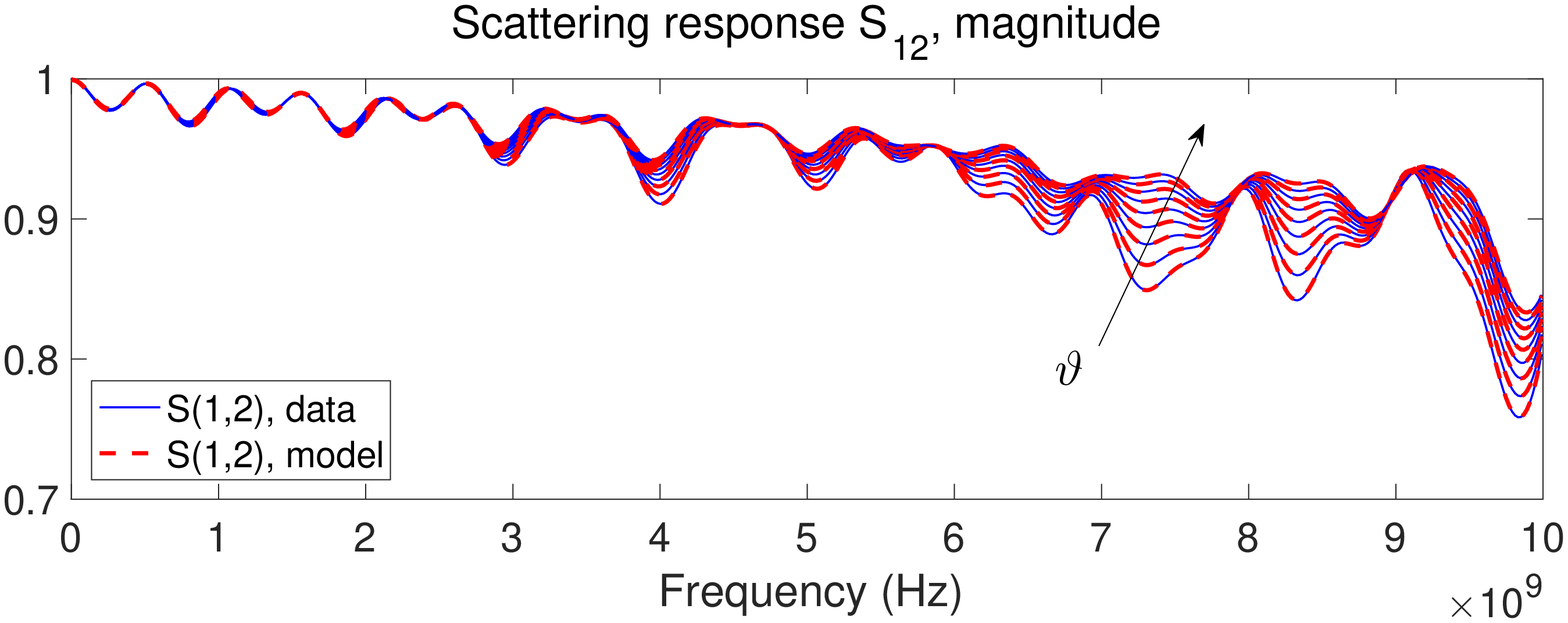}}
	\caption{Scattering responses $S_{ij}(\jj\omega;\vartheta)$ of the PCB link macromodel after passivity enforcement (dashed red lines), compared to the raw data used for model identification (solid blue lines). Both fitting and validation points are displayed; the arrows denote the increasing direction for via antipad radius $\vartheta$.}
	\label{fig:Slink_accuracy}
\end{figure}

\begin{figure}
	\centerline{\includegraphics[width=\columnwidth,clip]{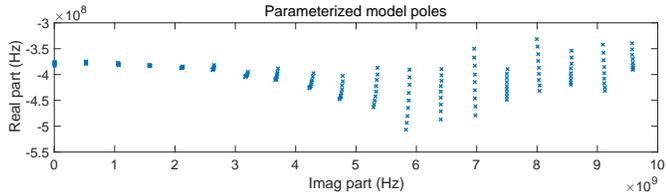}}
	\caption{Parameterized poles (zoomed view) of the PCB link macromodel, computed and superimposed for all original parameter samples $\vartheta_m$.}
	\label{fig:Slink_poles}
\end{figure}

The proposed passivity enforcement algorithm required three iterations and a runtime of 20 seconds. The final model, as evident from Fig.~\ref{fig:Slink_psi_passive}, is uniformly passive since $\psi(\vartheta)>0$ for all $\vartheta \in \Theta$. Figure~\ref{fig:Slink_accuracy} shows a comparison between the model responses and the raw scattering samples for all fitting and validation points. We see that the accuracy is excellent, as confirmed by the worst-case RMS errors $8.792 \times 10^{-4}$ (absolute) and $3.368 \times 10^{-3}$ (relative). These errors are only marginally worse than the errors of the original non-passive model, thanks to the adopted cost function~\eqref{eq:cost_function} that is minimized through passivity enforcement. This level of accuracy can be achieved only when the original passivity violations are very small (the worst-case passivity violation corresponded to a maximum singular value $\sigma_{\max}=1.000346$, which was larger than one by a very small amount). Finally, we depict the resulting implicitly-parameterized macromodel poles in Fig.~\ref{fig:Slink_poles}, computed by instantiating the model at the original parameter samples $\vartheta_m$.

%%%%%%%%%%%%%%%%%%%%%%%%%%%%%%%%%%%%%%%%%%%%%%%%%%%%%%%%%%%
\subsection{An integrated inductor}\label{sec:inductor}
%%%%%%%%%%%%%%%%%%%%%%%%%%%%%%%%%%%%%%%%%%%%%%%%%%%%%%%%%%%

We consider here a 2-port integrated inductor (courtesy of Prof. Madhavan Swaminathan, Georga Institute of Technology, Atlanta, USA). The inductor has a square outline with 1.5 turns routed on two different layers of the substrate. The scattering responses were obtained through a full-wave field solver for $\bar{m}=11$ different values of the sidelength $L=\vartheta$ ranging from~1.02 to~1.52~mm, with $\bar{k}=477$ frequency samples up to 12~GHz. Model identification via GSK iteration was performed using only odd-indexed parameter values, leaving the even-numbered responses for model validation purposes, with $\bar{n}=8$ poles and $\bar{l}=4,3$ Chebychev polynomial basis functions for numerator and denominator, respectively.

\begin{figure}
	\centerline{\includegraphics[width=\columnwidth,clip]{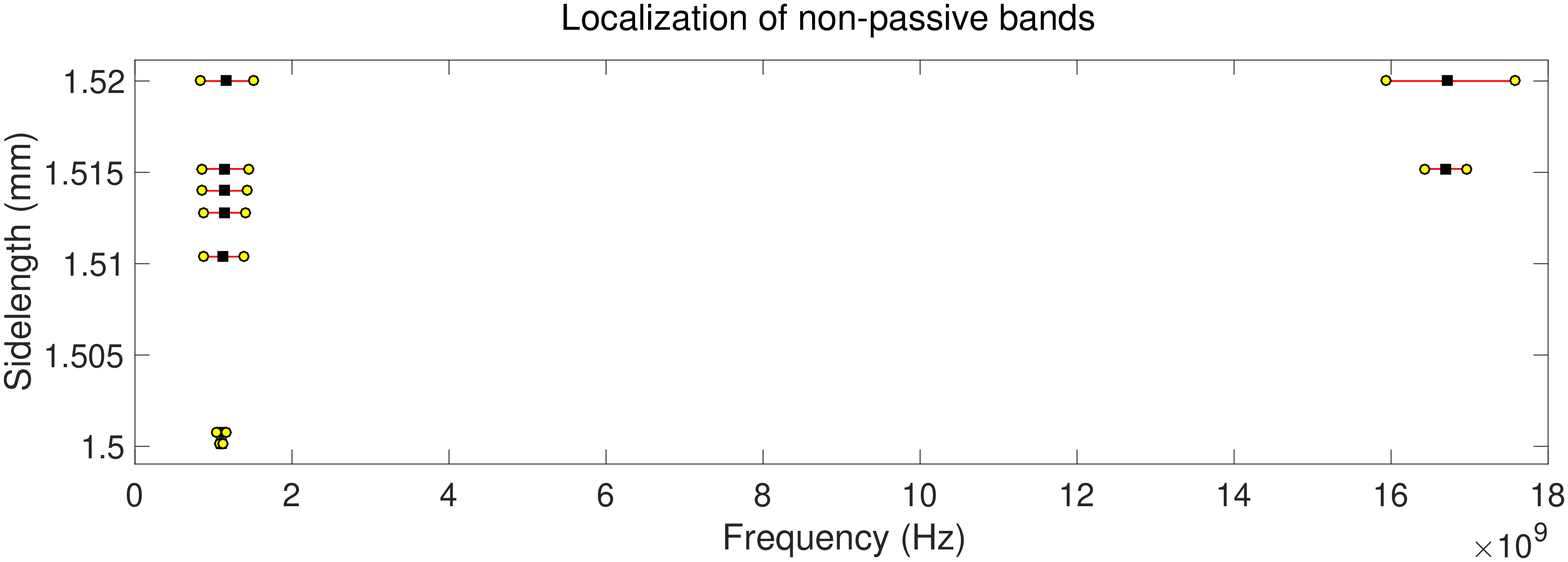}}
	\centerline{\includegraphics[width=\columnwidth,clip]{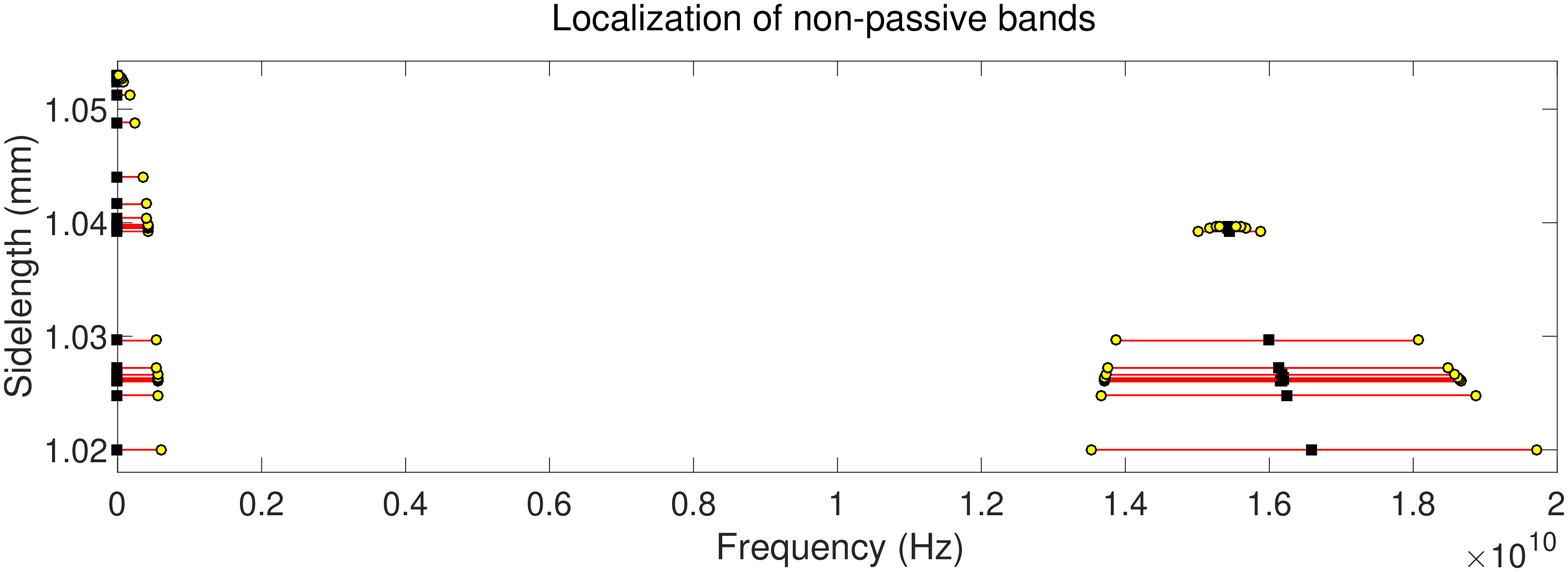}}
	\caption{Localization of non-passive bands (red lines) through imaginary SHH eigenvalues (yellow dots) of original non-passive inductor model. Top and bottom panels zoom on two different regions of the frequency-parameter plane.}
	\label{fig:inductor_passivity}
\end{figure}

\begin{figure}
	\centerline{\includegraphics[width=\columnwidth,clip]{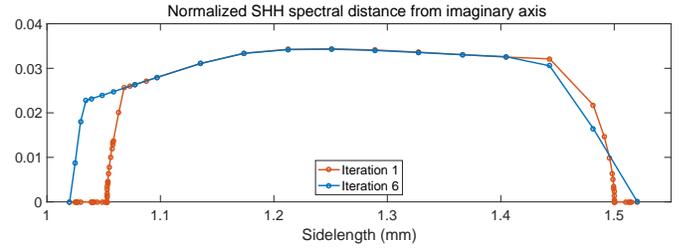}}
	\caption{Normalized SHH spectral distance from imaginary axis for the integrated inductor example, before (Iteration 1) and after (Iteration 6) passivity enforcement.}
	\label{fig:inductor_psi}
\end{figure}

\begin{figure}
	\centerline{\includegraphics[width=\columnwidth,clip]{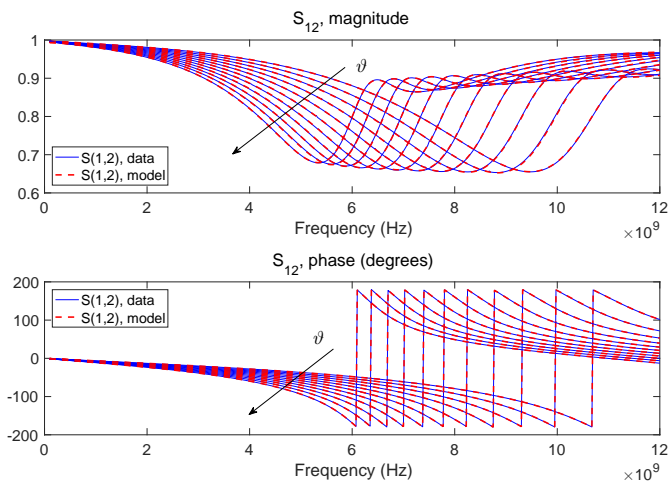}}
	\caption{Scattering response $S_{12}(\jj\omega;\vartheta)$ of the integrated inductor macromodel after passivity enforcement (dashed red lines), compared to the raw data used for model identification (solid blue lines). Both fitting and validation points are displayed; the arrows denote the increasing direction for sidelength $\vartheta$.}
	\label{fig:inductor_accuracy}
\end{figure}

The initial model was characterized by various passivity violations, illustrated in Fig.~\ref{fig:inductor_passivity}. These violations required a total of 5 iterations (runtime 15 seconds) to be removed, resulting in a uniformly passive model throughout the parameter range. Removal of passivity violations is confirmed by Fig.~\ref{fig:inductor_psi}, which depicts $\psi(\vartheta)$ before and after passivity enforcement. The RMS errors of original and passive models with respect to the original scattering responses are reported in Table~\ref{tab:inductor}, showing that both models are very accurate. Figure~\ref{fig:inductor_accuracy} confirms this statement by comparing passive model responses to raw data for all fitting and validation points.

\begin{table}
\caption{Absolute and relative RMS errors of non-passive ($\model{\mat{H}}$) and passive ($\pert{\model{\mat{H}}}$) inductor models at fitting and validation points.}
\label{tab:inductor}
\begin{tabular}{|c|c|c|c|c|}
\hline
Error & \multicolumn{2}{c|}{Validation points} & \multicolumn{2}{c|}{Fitting points} \\
\cline{2-5}
&             abs  &      rel &           abs &   rel  \\
\hline
$\model{H}_{11}$  & $1.84\times 10^{-3}$  & $4.64\times 10^{-3}$   &   $7.09\times 10^{-4}$  & $1.90\times 10^{-3}$ \\
$\model{H}_{12}$   & $2.60\times 10^{-3}$  & $2.96\times 10^{-3}$   &   $1.74\times 10^{-3}$  & $1.95\times 10^{-3}$ \\
$\model{H}_{22}$  & $2.04\times 10^{-3}$  & $5.13\times 10^{-3}$   &   $8.91\times 10^{-4}$  & $2.24\times 10^{-3}$ \\
\hline
$\pert{\model{H}}_{11}$  & $1.80\times 10^{-3}$  & $4.54\times 10^{-3}$   &   $8.58\times 10^{-4}$  & $2.12\times 10^{-3}$ \\
$\pert{\model{H}}_{12}$  & $2.56\times 10^{-3}$  & $2.92\times 10^{-3}$   &   $1.72\times 10^{-3}$  & $1.93\times 10^{-3}$ \\
$\pert{\model{H}}_{22}$  & $1.94\times 10^{-3}$  & $4.87\times 10^{-3}$   &   $1.10\times 10^{-3}$  & $2.72\times 10^{-3}$ \\
\hline
\end{tabular}
\end{table}

%%%%%%%%%%%%%%%%%%%%%%%%%%%%%%%%%%%%%%%%%%%%%%%%%%%%%%%%%%%
\subsection{A filter}\label{sec:filter}
%%%%%%%%%%%%%%%%%%%%%%%%%%%%%%%%%%%%%%%%%%%%%%%%%%%%%%%%%%%

\begin{figure}
	\centerline{\includegraphics[width=\columnwidth,clip]{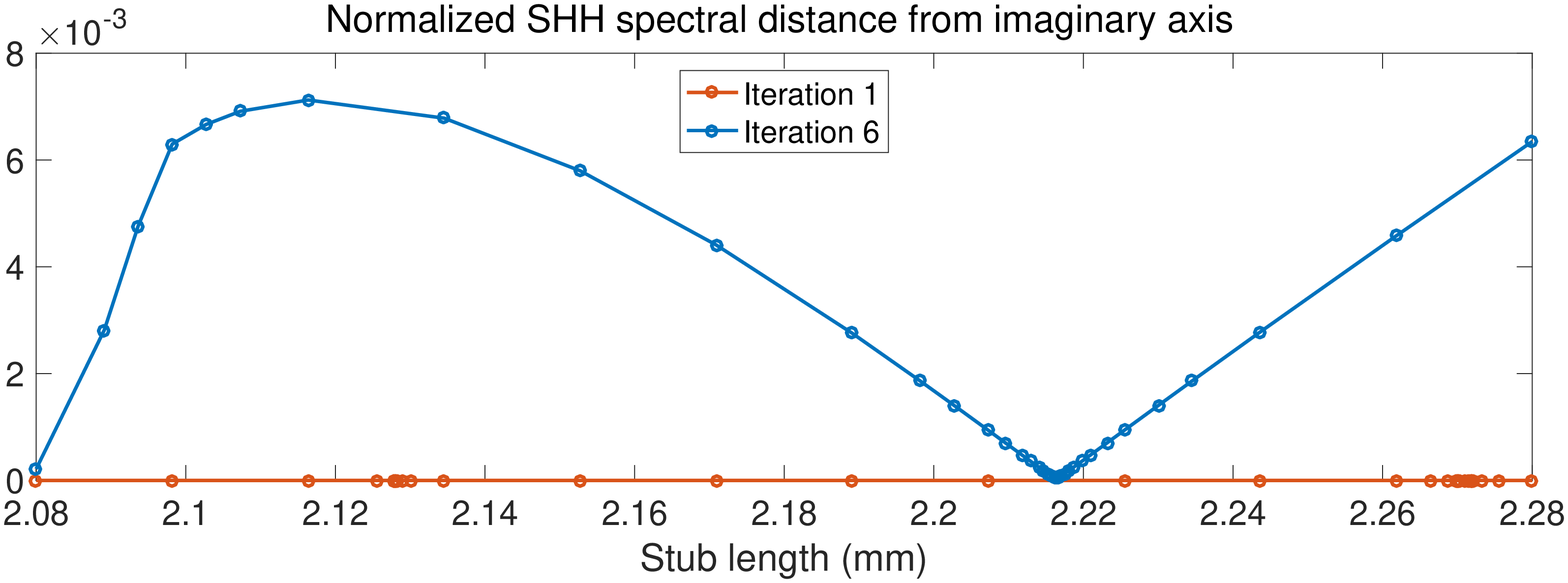}}
	\caption{Normalized SHH spectral distance from imaginary axis for the double-folded microstrip filter example, before (Iteration 1) and after (Iteration 6) passivity enforcement.}
	\label{fig:filter_psi}
\end{figure}

\begin{figure}
	\centerline{\includegraphics[width=\columnwidth,clip]{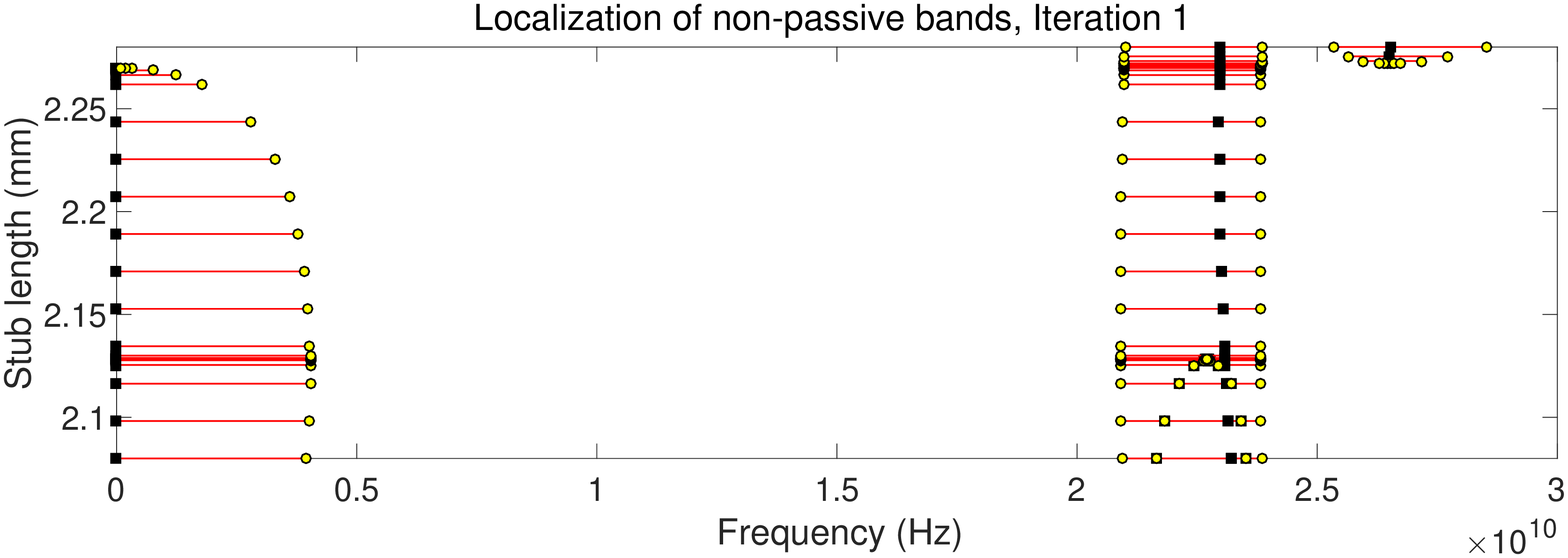}}
	\centerline{\includegraphics[width=\columnwidth,clip]{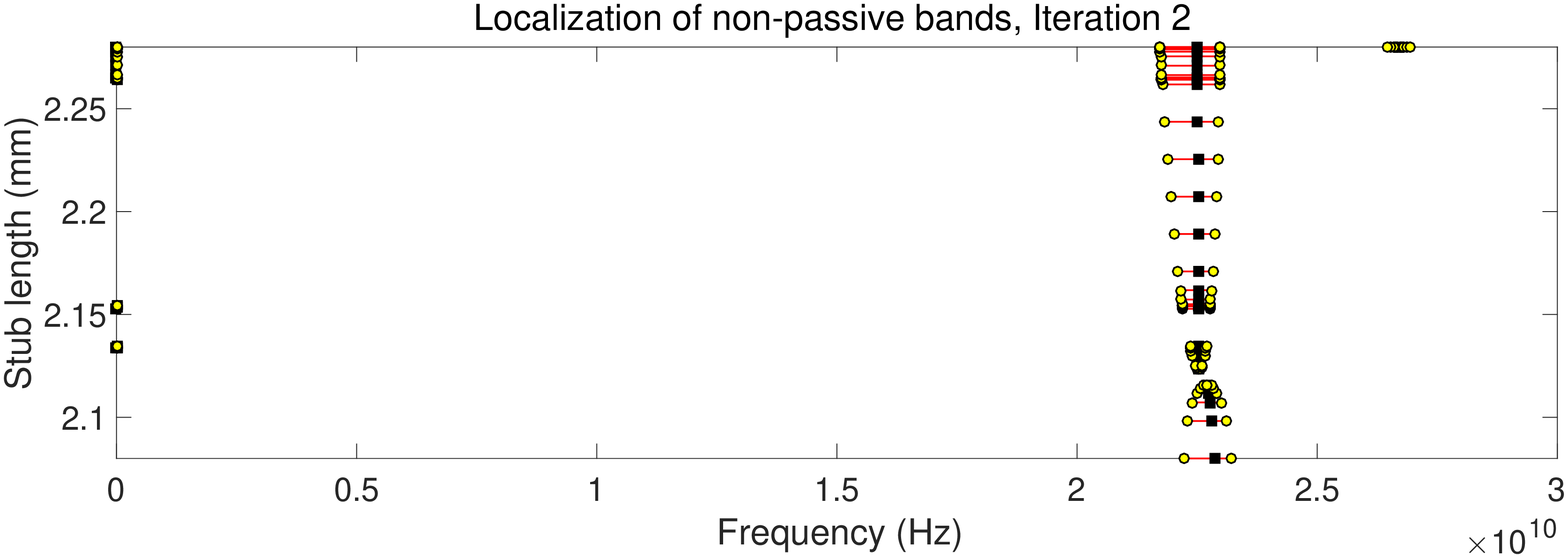}}
	\caption{Localization of non-passive bands (red lines) through imaginary SHH eigenvalues (yellow dots) of filter model at iteration~1 (top) and~2 (bottom) of the passivity enforcement loop.}
	\label{fig:filter_passivity}
\end{figure}

The last example we propose is a double-folded microstrip filter (see~\cite{cnf-2010-eumw-passiveparametricS} for a more detailed description), which can be tuned by changing the length $\vartheta=L$ of a microstrip stub within the range 2.08--2.28~mm. Scattering responses for $\bar{m}=21$ linearly-spaced stub length values were computed through a field solver at $\bar{k}=300$ samples spanning the frequency band $[5,20]$~GHz. Model identification based on odd-indexed parameter samples required $\bar{n}=10$ poles and $\bar{l}=3$ (quadratic) polynomial basis functions to capture parameter variations.

The resulting model was not passive, as Fig.~\ref{fig:filter_psi} shows by depicting $\psi(\vartheta)=0$ at the first passivity iteration loop. The localized passivity violations of the initial model are depicted in the top panel of Fig.~\ref{fig:filter_passivity}. Passivity enforcement required 5 iterations (runtime 31 seconds) to achieve uniform passivity, as confirmed by $\psi(\vartheta)>0$ in Fig.~\ref{fig:filter_psi}. The passivity violations at the second iteration are depicted in the bottom panel of Fig.~\ref{fig:filter_passivity}, where we can note that the extent of such violations both in the frequency and parameter directions is reduced. Note also that the presence of multiple singular values exceeding one (Fig.~\ref{fig:filter_passivity}, Iteration 1, localized at 22--24~GHz and approximately $\vartheta\in[2.08,2.13]$~mm), does not pose particular problems, since multiple independent constraints can be enforced while solving~\eqref{eq:single_iteration}.

\begin{figure}
	\centerline{\includegraphics[width=\columnwidth,clip]{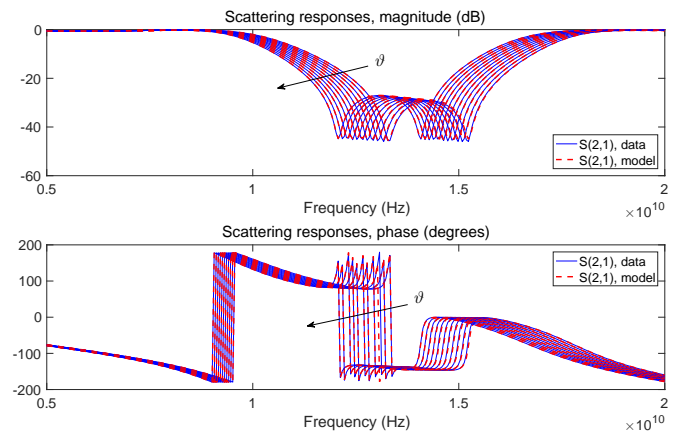}}
	\caption{Scattering responses $S_{21}(\jj\omega;\vartheta)$ of the filter macromodel after passivity enforcement (dashed red lines), compared to the raw data used for model identification (solid blue lines). Only validation points are displayed; the arrows denote the increasing direction for stub length $\vartheta$.}
	\label{fig:filter_accuracy}
\end{figure}

Finally, Fig.~\ref{fig:filter_accuracy} compares the model responses to the original scattering responses at the validation points (even-indexed parameter samples). Also for this example the accuracy is excellent, with a worst-case RMS error amont all parameter, frequency samples and scattering matrix elements equal to $3.16\times 10^{-3}$ (absolute) and $4.64\times 10^{-3}$ (relative).

%%%%%%%%%%%%%%%%%%%%%%%%%%%%%%%%%%%%%%%%%%%%%%%%%%%%%%%%%%%
\section{Conclusion}\label{sec:conclusion}
%%%%%%%%%%%%%%%%%%%%%%%%%%%%%%%%%%%%%%%%%%%%%%%%%%%%%%%%%%%

This paper presented an efficient and robust algorithm for checking and enforcing the passivity of behavioral macromodels of LTI systems, whose scattering matrix depends both on frequency and on one additional external parameter. Thanks to a specialized adaptive sampling process in the parameter space, based on the spectral properties of some Skew-Hamiltonian/Hamiltonian matrix pencil associated to the model, we are able to detect and localize all frequency- and parameter-dependent passivity violations of the model. An iterative constrained optimization loop is able to remove such violations, while retaining model accuracy. To be best of Author's knowledge, this is the first documented algorithm that is able to enforce the uniform passivity of a parameterized macromodel.

Future research will be devoted to the extension of proposed approach to higher dimensions in the parameter space, with special emphasis on avoiding issues due to the curse of dimensionality.

%%%%%%%%%%%%%%%%%%%%%%%%%%%%%%%%%%%%%%%%%%%%%%%%%%%%%%%%%%%
\section{Acknowledgements}\label{sec:ack}
%%%%%%%%%%%%%%%%%%%%%%%%%%%%%%%%%%%%%%%%%%%%%%%%%%%%%%%%%%%

The Author is grateful to Prof.~Madhavan Swaminathan (Georgia Institute of Technology, Atlanta, USA) for sharing the integrated inductor data, to Prof. Christian Schuster and Dr. Jan Preibisch (Technische Universit\"at Hamburg-Harburg, Hamburg, Germany) for sharing the PCB interconnect link data, and to Prof. Piero Triverio (Univ. Toronto, Canada) for sharing the filter data.

%%%%%%%%%%%%%%%%%%%%%%%%%%%%%%%%%%%%%%%%%%%%%%%%%%%%%%%%%%%

%%%%%%%%%%%%%%%%%%%%%%%%%%%%%%%%%%%%%%%%%%%%%%%%%%%%%%%%%%%


\begin{thebibliography}{99}

\bibitem{PM_book}
S.~Grivet-Talocia and B.~Gustavsen, \emph{Passive Macromodeling: Theory and Applications}. New York: John Wiley and Sons, 2016.

\bibitem{jnl-2010-temc-pi}
M. Swaminathan, D. Chung, S. Grivet-Talocia, K. Bharath, V. Laddha, and J. Xie, ``Designing and modeling for power integrity,'' IEEE Transactions on Electromagnetic Compatibility, vol. 52, pp. 288Ð310, May 2010.

\bibitem{antoulas2005}
A.C. Antoulas, {\em Approximation of large-scale dynamical systems}, SIAM, Philadelphia, 2005.

\bibitem{schilders2008}
W.H.A. Schilders, H.A. Van Der~Vorst, and J.~Rommes, {\em Model order reduction: theory, research aspects and applications}, Springer Verlag, 2008.

\bibitem{nakhla01}
M.~Nakhla and R.~Achar,
``Simulation of High-Speed Interconnects'', {\em Proc. IEEE}, May 2001,
Vol.~89, No.~5, pp.~693--728.

\bibitem{CelikBook} M.~Celik, L.~Pileggi, A.~Odabasioglu, \emph{IC Interconnect Analysis}, Springer, 2002.

\bibitem{Gustavsen99} B. Gustavsen, A. Semlyen, ``Rational approximation of frequency domain responses by vector fitting'', \emph{IEEE Trans. Power Del.}, vol. 14, no. 3, pp. 1052-1061, July, 1999.

\bibitem{ovf2}
D.~Deschrijver, B.~Haegeman, T.~Dhaene,
``Orthonormal Vector Fitting: A Robust Macromodeling Tool for Rational Approximation of Frequency Domain Responses,'' {\em IEEE Trans. Adv. Packaging}, vol.~30, pp.~216--225, May 2007.

\bibitem{desc2008}
D.~Deschrijver, M.~Mrozowski, T.~Dhaene, D.~De~Zutter, ``Macromodeling of Multiport Systems Using a Fast Implementation of the Vector Fitting Method,'' {\em IEEE Microwave and Wireless Components Letters}, Vol.~18, N.~6, June 2008, pp.383--385.

%\bibitem{tan2006}
%Z. Qi, H. Yu, P. Liu, S.X.Tan, L. He, ``Wideband passive multiport model order reduction and realization of RLCM circuits,'' \emph{IEEE Trans. on Computer-Aided Design of Integrated Circuits and Systems}, vol.~25, no.~8, pp.~1496--1509, Aug. 2006

%\bibitem{antonini2003}
%G.~Antonini, ``SPICE equivalent circuits of frequency-domain responses,'' \emph{IEEE Trans. on Electromagnetic Compatibility}, vol.~45, no.~3, pp.~502--512, Aug. 2003

\bibitem{artAJMACA}
A.~J. Mayo and A.~C. Antoulas, ``A framework for the solution of the
  generalized realization problem,'' \emph{Linear Algebra and Its
  Applications}, vol. 405, no. 2-3, pp. 634--662, 2007.

\bibitem{SLACATCAD09}
S.~Lefteriu and A.~C. Antoulas, ``A new approach to modeling multi-port systems
  from frequency domain data,'' vol.~29, no.~1, pp. 14 --27, Jan. 2010.

\bibitem{Anderson}
B.~D.~O. Anderson and S.~Vongpanitlerd.
\newblock {\em Network analysis and syntesis}.
\newblock Prentice-Hall, 1973.

\bibitem{Wohlers}
M.~R. Wohlers.
\newblock {\em Lumped and Distributed Passive Networks}.
\newblock Academic press, 1969.

\bibitem{Triverio07} P. Triverio, S. Grivet-Talocia, M.S. Nakhla, F. Canavero, R. Achar, ``Stability, causality, and passivity in electrical interconnect models'', \emph{IEEE Trans. on Advanced Packaging}, vol. 30, no. 4, pp. 795-808, 2007.

\bibitem{jnl-2009-ijcta-destabilize}
S. Grivet-Talocia, ``On driving non-passive macromodels to instability,'' International Journal of Circuit Theory And Applications, vol. 37, pp. 863Ð886, Oct 2009.

\bibitem{BBK89}
S.~Boyd, V.~Balakrishnan, P.~Kabamba,
``A bisection method for computing the $H_\infty$ norm of a
transfer matrix and related problems'',
{\em Math. Control Signals Systems}, Vol.~2, 1989, pp.~207--219.

\bibitem{LMI}
S.~Boyd, L.~El Ghaoui, E.~Feron, V.~Balakrishnan,
{\em Linear matrix inequalities in system and control theory},
{\em SIAM studies in applied mathematics}, SIAM, Philadelphia, 1994.

\bibitem{NWCKC2008}
N.~Wong, C.-K.~Chu, ``A fast passivity test for stable descriptor systems via skew-Hamiltonian/Hamiltonian matrix pencil transformations.'' IEEE Transactions on Circuits and Systems I: Regular Papers vol.~55, no.~2, 2008, pp.~635-643.

\bibitem{ZZNW2010a}
Z. Zhang and N. Wong, ``Passivity Test of Immittance Descriptor Systems Based on Generalized Hamiltonian Methods,'' in IEEE Transactions on Circuits and Systems II: Express Briefs, vol. 57, no. 1, pp. 61-65, Jan. 2010.

\bibitem{ZZNW2010b}
Z. Zhang and N. Wong, ``An Efficient Projector-Based Passivity Test for Descriptor Systems,'' in IEEE Transactions on Computer-Aided Design of Integrated Circuits and Systems, vol. 29, no. 8, pp. 1203-1214, Aug. 2010.

\bibitem{ZZNW2010c}
Z. Zhang and N. Wong, ``Passivity Check of  S -Parameter Descriptor Systems via  S-Parameter Generalized Hamiltonian Methods,'' in IEEE Transactions on Advanced Packaging, vol. 33, no. 4, pp. 1034-1042, Nov. 2010.

\bibitem{SHH_perturbation}
Y. Wang, Z. Zhang, C. K. Koh, G. Shi, G. K. H. Pang and N. Wong, ``Passivity Enforcement for Descriptor Systems Via Matrix Pencil Perturbation,'' in IEEE Transactions on Computer-Aided Design of Integrated Circuits and Systems, vol. 31, no. 4, pp. 532-545, April 2012.

\bibitem{Grivet04pass} S. Grivet-Talocia, ``Passivity Enforcement via Perturbation of Hamiltonian Matrices" , in \emph{IEEE Trans. Circuits and Systems I: Fundamental Theory and Applications}, pp. 1755-1769, vol. 51, n. 9, September, 2004.

\bibitem{saraswat2005}
D.~Saraswat, R.~Achar and M.~Nakhla, ``Global Passivity Enforcement Algorithm for Macromodels of Interconnect Subnetworks Characterized by Tabulated Data'', {\em IEEE Transactions on VLSI Systems}, Vol.~13, No.~7, pp. 819--832, July~2005.

\bibitem{Gust01}
B.~Gustavsen, A.~Semlyen,
``Enforcing passivity for admittance matrices approximated by rational functions'',
{\em IEEE Trans. Power Systems}, Vol.~16, N.~1, pp.~97--104, Feb.~2001.

\bibitem{jnl-2008-tadvp-PassivityMethods}
S.~Grivet-Talocia and A.~Ubolli, ``A comparative study of passivity enforcement
  schemes for linear lumped macromodels,'' {\em {IEEE} Trans. Advanced
  Packaging}, vol.~31, pp.~673--683, Nov 2008.

\bibitem{silveira}{
C. P. Coelho, J. Phillips, and L. M. Silveira, ``A Convex Programming Approach for
Generating Guaranteed Passive Approximations to Tabulated Frequency-Data,''
{\em IEEE Trans. Computed-Aided Design of Integrated Circuits and Systems},
vol.~23, no.~2, pp.~293--301, Feb.~2004.}

\bibitem{jnl-2012-tmtt-subgradient}
G. Calafiore, A. Chinea, and S. Grivet-Talocia, ``Subgradient techniques for passivity enforcement of linear device and inter- connect macromodels,'' IEEE Transactions on Microwave Theory and Techniques, vol. 60, pp. 2990Ð3003, October 2012.

\bibitem{jnl-2014-tcad-localization}
Z. Mahmood, S. Grivet-Talocia, A. Chinea, G. Calafiore, and L. Daniel, ``Efficient localization methods for passivity enforcement of linear dynamical models,'' IEEE Transactions on Computer-Aided Design of Integrated Circuits and Systems, vol. 33, pp. 1328Ð1341, Sept 2014.

\bibitem{TriverioPhD}
P. Triverio, ``Self Consistent, Efficient and Parametric Macromodels for High-speed Interconnects Design", Ph.D. Thesis, Politecnico di Torino, Italy, April 2009.

\bibitem{jnl-2007-temc-Parameterization}
S.~Grivet-Talocia, S.~Acquadro, M.~Bandinu, F.~G. Canavero, I.~Kelander, and
  M.~Rouvala, ``A parameterization scheme for lossy transmission line
  macromodels with application to high speed interconnects in mobile devices,''
  {\em {IEEE} Trans. Electromagnetic Compatibility}, vol.~49, no.~1, pp.~18--24,
  Feb.~2007.

\bibitem{Triverio2009}
P. Triverio, S. Grivet-Talocia, and M.S. Nakhla, ``A Parameterized Macromodeling Strategy with Uniform Stability Test,'' \emph{IEEE Transactions on Advanced Packaging}, vol.~32, no.~1, pp.~205-215, Feb.~2009.

\bibitem{Triverio10}
P. Triverio, M.Nakhla and S. Grivet-Talocia, ``Passive Parametric Modeling of Interconnects and Packaging Components from Sampled Impedance, Admittance or Scattering Data,'' \emph{Electronics System Integration Technology Conferences (ESTC)}, Berlin (Germany), Sept.~13--16, 2010, pp. 1-6.

\bibitem{Triverio07EPEP}
P. Triverio, M. Nakhla, and S. Grivet-Talocia, ``Parametric Macromodeling of Multiport Networks from Tabulated Data,'' \emph{IEEE 16th Topical Meeting on Electrical Performance of Electronic Packaging (EPEP 2007)}, Atlanta (GA), USA, Oct.~29--31, 2007, pp.~51--54.

\bibitem{jnl-2014-tcpmt-small-signal}
S.~B.~Olivadese, G.~Signorini, S.~Grivet-Talocia, P.~Brenner, ``Parameterized and DC-compliant small-signal
macromodels of RF circuit blocks,'' \emph{IEEE Transactions on Components, Packaging, and Manufacturing Technology}, vol.~5, pp.~508-Ð522, April 2015.

\bibitem{jnl-2017-temc-fourier-rational}
S. Grivet-Talocia and E. Fevola, "Compact Parameterized Black-Box Modeling via Fourier-Rational Approximations," in IEEE Transactions on Electromagnetic Compatibility, vol. 59, no. 4, pp. 1133-1142, Aug. 2017.

\bibitem{Samuel13}
E. R. Samuel, L. Knockaert, F. Ferranti, and T. Dhaene, ``Guaranteed Passive Parameterized Macromodeling by Using Sylvester State-Space Realizations", \emph{IEEE Transactions on Microwave Theory and Techniques}, vol. 61, no. 4, pp. 1444-1454, Mar.~2013.

\bibitem{Ferranti12b}{
F.~Ferranti, T.~Dhaene, and L.~Knockaert, ``Compact and Passive Parametric Macromodeling Using Reference Macromodels and Positive Interpolation Operators,'' \emph{IEEE Transactions on Components, Packaging and Manufacturing Technology}, vol.~2, no.~12, pp.~2080--2088, Dec.~2012.}

\bibitem{Ferranti11}{
F.~Ferranti, L.~Knockaert, and T.~Dhaene, ``Passivity-Preserving Parametric Macromodeling by Means of Scaled and Shifted State-Space Systems,'' \emph{IEEE Transactions on Microwave Theory and Techniques}, vol.~59, no.~10, pp.~2394--2403, Oct.~2011.}

\bibitem{Ferranti10b}
F.~Ferranti, L.~Knockaert, T.~Dhaene, ``Guaranteed Passive Parameterized Admittance-Based Macromodeling,'' \emph{IEEE Transactions on Advanced Packaging}, vol.~33, no.~3, pp.~623--629, Aug.~2010.

\bibitem{San63}{
C. K. Sanathanan and J. Koerner, ``Transfer function synthesis as a ratio of two complex polynomials,'' \emph{{IEEE} Trans. Automatic Control}, vol.~8, no.~1, pp.~56--58, Jan.~1963.}

\bibitem{Chinea2011}
A. Chinea, S. Grivet-Talocia, H. Hu, P. Triverio, D. Kaller, C. Siviero, and M. Kindscher, ÒSignal integrity verification of multichip links using passive channel macromodels,Ó {\em IEEE Transactions on Components, Packaging, and Manufacturing Technology}, vol. 1, pp. 920Ð933, June 2011

\bibitem{chiang05}
I.-T. Chiang et al., ``Fast Real-Time Convolution Algorithm for Microwave Multiport Networks With Nonlinear Terminations,'' {\em IEEE Transactions on Circuits and Systems -- II: Express Briefs}, Vol. 52 (2005).

\bibitem{lefteriu2013}
S. Lefteriu and A. C. Antoulas, "On the Convergence of the Vector-Fitting Algorithm," in {\em IEEE Transactions on Microwave Theory and Techniques}, vol. 61, no. 4, pp. 1435-1443, April 2013.

\bibitem{shi2016}
G. Shi, "On the Nonconvergence of the Vector Fitting Algorithm," in {\em IEEE Transactions on Circuits and Systems II: Express Briefs}, vol. 63, no. 8, pp. 718-722, Aug. 2016.

\bibitem{SHHlarge}
V.~Mehrmann, D.~Watkins, ``Structure-preserving methods for computing eigenpairs of large sparse skew-Hamiltonian/Hamiltonian pencils.'' SIAM Journal on Scientific Computing, vol.~22, no.~6, 2001, pp.~1905-1925.

\bibitem{Watkins2004}
D.S.~Watkins, ``On Hamiltonian and symplectic Lanczos processes.'' Linear algebra and its applications, vol.~385, 2004, pp.~23-45.

\bibitem{benner2002}
P.~Benner, R.~Byers, V.~Mehrmann, H.~Xu, ``Numerical computation of deflating subspaces of skew-Hamiltonian/Hamiltonian pencils,'' SIAM Journal on Matrix Analysis and Applications, vol.~24, no.~1, pp.~165-190, 2002.

\bibitem{bai}
Z. Bai, J. Demmel, J. Dongarra, A. Ruhe and H. van der Vorst, editors, \emph{Templates for the solution of Algebraic Eigenvalue Problems: A Practical Guide.} SIAM, Philadelphia, 2000

\bibitem{boyd2004}
S.~P. Boyd and L.~Vandenberghe.
\newblock {\em Convex optimization}.
\newblock Cambridge Univ Pr, 2004.

\bibitem{SchusterDesignCon2017}
J.~B.~Preibisch, T.~Reuschel, K.~Scharff, J.~Balachandran, B.~Sen, C.~Schuster, ``Exploring Efficient Variability-Aware Analysis Method for High-Speed Digital Link Design Using PCE'', DesignCon, Jan 31-Feb 2, 2017, Santa Clara (CA), USA.

\bibitem{CIM}
X. Duan, R. Rimolo-Donadio, H.-D. Br\"uns, and C. Schuster, ``Circular ports in parallel-plate waveguide analysis with isotropic excitations,'' {\em IEEE Transactions on Electromagnetic Compatibility}, vol. 54, pp. 603Ð612, June 2012.

\bibitem{cnf-2010-eumw-passiveparametricS}
P. Triverio, M. Nakhla, and S. Grivet-Talocia, ÒExtraction of parametric circuit models from scattering parameters of passive RF components,Ó in Proc. of the 5th European Microwave Integrated Circuits Conference, (Paris), pp. 393Ð396, September 27-28 2010.

\end{thebibliography}
\end{document}